\pdfoutput=1
\documentclass[11pt]{article}

\usepackage[T1]{fontenc}
\usepackage[utf8]{inputenc}
\usepackage{lmodern}

\usepackage[margin=1in]{geometry}
\usepackage{amsmath,amssymb,amsfonts,bm}
\usepackage{mathtools}

\usepackage{graphicx}
\graphicspath{{./}}
\usepackage{booktabs}
\usepackage{tikz}
\usetikzlibrary{decorations.pathmorphing,arrows.meta,positioning}

\usepackage{microtype}
\usepackage{xcolor}
\usepackage{slashed}
\usepackage{cite}
\usepackage{authblk}
\usepackage{float}
\usepackage[
    colorlinks=true,
    linkcolor=blue,
    citecolor=blue,
    urlcolor=blue,
    hyperfootnotes=true
]{hyperref}

\hypersetup{
    pdftitle={A Minimal Singlet-Scalar and Vector-Like Quark Benchmark for B -> Xs gamma and Bs Mixing},
    pdfauthor={Author Name}
}
\usepackage{booktabs}
\usepackage[table]{xcolor}
\usepackage{array}
\usepackage{caption}
\usepackage{tikz-feynman}
\usetikzlibrary{positioning}
\tikzfeynmanset{compat=1.1.0}

\usetikzlibrary{decorations.pathmorphing,arrows.meta}

\tikzset{
  fermion/.style={line width=0.75pt,-{Latex[length=1.8mm,width=1.3mm]}},
  scalar/.style={line width=0.75pt,dashed},
  photon/.style={
    line width=0.75pt,
    decorate,
    decoration={snake,amplitude=0.7mm,segment length=3.0mm}
  },
  gluon/.style={
    line width=0.75pt,
    decorate,
    decoration={coil,amplitude=0.9mm,segment length=2.2mm}
  },
  vtx/.style={circle,fill=black,inner sep=1.4pt},
  vf/.style={font=\scriptsize}
}
\numberwithin{equation}{section}

\title{\Large \bf Real and Complex Singlet-Scalar Benchmarks with a Vector-Like Down Quark for $B\to X_s\gamma$ and $B_s-\bar B_s$ Mixing}

\author[1]{Qazi Maaz Us Salam\footnote{{\color{blue}qazimaaz92@gmail.com}}}

\affil[1]{\small \textit{Department of Physics, Lahore University of Management Sciences (LUMS),
Opposite Sector U, D.H.A, Lahore 54792, Pakistan}}

\begin{document}
\date{}
\maketitle

\begin{abstract}
We study a simple extension of the Standard Model with a vector-like down-type quark $D$ and a neutral singlet scalar ${\cal S}=S_R,\Phi$. The scalar is considered in two forms, a real field \(S_R=S_R^\dagger\) and a complex field $\Phi\neq\Phi^\dagger$. The interaction $-\lambda_i{\cal S}\bar D_L d_{Ri}+{\rm h.c.}$ generates the radiative
transitions \(b\to s\gamma\) and \(b\to sg\) at one loop. Since ${\cal S}$ has no electric charge or color, the gauge boson is emitted from the internal $D$ line, giving $C_{7\gamma}^{\rm NP}/C_{8G}^{\rm NP}=Q_D=-1/3$ at the matching scale for $\Delta B=1$ dipole transitions. For $M_D=m_{\cal S}=1~{\rm TeV}$ and $|\lambda_s^\ast\lambda_b|=1$, the low scale contribution is
$|C_{7\gamma}^{\rm NP,eff}(\mu_b)|\simeq 1.1\times10^{-3}$, This is about $0.4\%$ of the Standard Model value
$|C_{7\gamma}^{\rm SM,eff}(\mu_b)|\simeq 0.30$. We also discuss \(B_s-\bar B_s\) mixing. In the real-scalar case, the direct and crossed box diagrams cancel in the exact minimal limit. In the complex-scalar case, the
direct box contribution remains and gives a bound on the flavor $|\lambda_s^\ast\lambda_b|$ at the level of a few tenths for TeV-scale masses. Thus, in these minimal benchmarks, $B\to X_s\gamma$ is radiatively safe, while $B_s-\bar B_s$ mixing gives the stronger constraint in the complex-scalar $\Phi$ benchmark.
\end{abstract}

\newpage

\section{Introduction}
The flavor changing neutral current (FCNC) processes do not occur at the tree level in the standard model of particle physics (SM) and appear at the loop level. For this reason, the leading order contributions to these processes can, in principle, receive corrections from physics beyond the SM. Therefore, FCNC processes provide an attractive theoretical and experimental tool for testing the SM and exploring any new physics (NP) or new interactions~\cite{London:2021lfn,Albrecht:2021tul,Salam:2024nfv,Aarfi:2025qcp,Aarfi:2026hgi}. In the SM, the quark-level FCNC transition $b\to s\gamma$ first appears at the loop level through electroweak penguin diagrams. This loop suppression makes the inclusive decay $B\to X_s\gamma$ particularly sensitive to new heavy particles~\cite{Haisch:2008ar,Borzumati:2003rr,Bernlochner:2020jlt,Adhikary:2025gdh,Hermann:2012fc}. These particles can appear in the loop and change the short-distance (SD) Wilson coefficients (WCs).

The effective Hamiltonian is a framework to separate SD physics from long-distance hadronic effects~\cite{Buchalla:1995vs,Buras:1998raa}. Further, we use this framework for radiative $B$ decays, the main contribution comes from the $O_{7\gamma}$ operator. The $O_{8G}$ operator is also important because, through QCD running, it mixes into $O_{7\gamma}$ when going from the high matching scale down to the low scale $\mu_b\sim m_b$. Therefore, a consistent NP study of $b\to s\gamma$ should include both the WCs, $C_{7\gamma}$ and $C_{8G}$.

In this context, vector-like fermions are one of the SM extensions~\cite{Alves:2023ufm, Banerjee:2024zvg,Barger:2007qb,Arhrib:2016rlj,Benbrik:2019zdp,CidVidal:2018eel}. The motivation behind this theory is the anomaly cancellation by itself. These particles can be called more specifically as standard vector-like particles. We call it standard because these vector-like particle transforms in the same way as SM fermions. If these fermions couple to SM quarks through an extra scalar field, they can contribute to rare FCNC processes at the loop level. In this work, we study the simplest version of this setup: a neutral gauge-singlet scalar ${\cal S}$ and a vector-like down-type quark $D$. The flavor transition $b\to s\gamma$ comes from the couplings $\lambda_s^\ast\lambda_b$. Since as the scalar is electrically neutral and a color singlet, the photon and gluon in the one-loop diagrams can only be emitted from the internal vector-like quark line. This gives a characteristic relation between the electromagnetic and chromomagnetic matching contributions.

Generic matching expressions for $b\to s\gamma$ and $b\to sg$ in extensions of the SM have been studied in detail by C. Bobeth \textit{et al}~\cite{Bobeth:1999ww}, these studies also include heavy fermion and heavy scalar loop effects together with their QCD corrections. More recent studies have also analyzed models with vector-like fermions and singlet scalars loop effects in $b\to s\ell^+\ell^-$, $B_s-\bar B_s$ mixing, and related observables~\cite{Grinstein:2018fgb,Arnan:2019uhr,Banerjee:2022xmu}. Furthermore, the phenomenological implications have also been reviewed in detail by~\cite{Alves:2023ufm}. Besides this, current collider searches increasingly emphasize that additional scalar decay channels can significantly modify the usual LHC bounds on vector-like quarks~\cite{Banerjee:2024zvg,Buchkremer:2013bha,ATLAS:2024itc}.

The present work focuses on benchmark analysis and does not perform a comprehensive global analysis of vector-like quark phenomenology. We analyze a minimal extension of the SM containing a neutral gauge-singlet scalar ${\cal S}$ and a vector-like down-type quark $D$. Here, the ${\cal S}$  is catagorized into two benchmarks: a real singlet scalar $S_R=S_R^\dagger$ and a complex singlet scalar $\Phi\neq\Phi^\dagger$. The new Yukawa-like interaction generates FCNC effects through the coupling product $\lambda_s^\ast\lambda_b$. We study the $B\to X_s\gamma$ corrections, and notice the absence of a chirality-enhanced $M_D/m_b$ contribution in the minimal coupling structure. This feature makes the benchmark radiatively safe. At the same time, it shows that other observables, especially ($\Delta B=2$) FCNC processes, can provide more restrictive limits in a full model-level analysis.

We now give the structure of the rest of this paper. In Section~\ref{sec:model}, we introduce the real and complex singlet-scalar with the vector-like quark model. In Section~\ref{sec:hamiltonian}, we review the effective Hamiltonian. In Section~\ref{sec:matching}, we compute the one-loop matching contributions to $C_{7\gamma}$ and $C_{8G}$. In Section~\ref{sec:mixing}, we discuss $B_s-\bar B_s$ mixing. In Section~\ref{sec:numerics}, we present the numerical analysis, including both the radiative bound and the associated $B_s-\bar B_s$ mixing constraint. Finally, in Section~\ref{sec:conclusion}, we present the conclusions and collect the supporting mathematical details in the appendices.

\section{Real and Complex Singlet-Scalar with Vector-Like Quark Model}
\label{sec:model}

We extend the SM by adding a vector-like down-type quark $D$. This follows the standard use of vector-like fermions as anomaly safe SM extensions~\cite{Alves:2023ufm,Banerjee:2024zvg}. Its quantum numbers under the SM gauge group are
\begin{equation}
D_{L,R}\sim (3,1,-1/3),
\end{equation}
where the entries denote representations under $SU(3)_c\times SU(2)_L\times U(1)_Y$. We then compare two minimal choices for the scalar sector:
\begin{equation}
S_R=S_R^\dagger\sim(1,1,0),
\qquad
\Phi\sim(1,1,0),\qquad \Phi\neq\Phi^\dagger.
\end{equation}
Here, $S_R$ denotes the real scalar, while $\Phi$ is the complex scalar field.
The relevant Lagrangian can be written as
\begin{align}
\mathcal L_R
&=
\mathcal L_{\rm SM}
+\bar D(i\slashed D-M_D)D
+\frac12(\partial_\mu S_R)(\partial^\mu S_R)
-\frac12m_S^2S_R^2
+\mathcal L_{\rm int}^{R},
\\
\mathcal L_C
&=
\mathcal L_{\rm SM}
+\bar D(i\slashed D-M_D)D
+(\partial_\mu\Phi)^\dagger(\partial^\mu\Phi)
-m_S^2\Phi^\dagger\Phi
+\mathcal L_{\rm int}^{C},
\end{align}
where $M_D$ is the vector-like quark mass and $m_S$ is the scalar singlet mass. The interaction terms relevant for $b\to s\gamma$ are chosen as a minimal right-handed scalar fermion coupling, in the same class of scalar fermion interactions used in generic loop analyses are given in~\cite{Grinstein:2018fgb,Arnan:2019uhr}:
\begin{align}
\mathcal L_{\rm int}^{R}
&=
-\lambda_i^R S_R\bar D_L d_{Ri}+{\rm h.c.},
\\
\mathcal L_{\rm int}^{C}
&=
-\lambda_i^C \Phi\bar D_L d_{Ri}+{\rm h.c.},
\qquad i=s,b.
\end{align}
For the radiative matching, both cases are described by the compact notation
\begin{equation}
\mathcal L_{\rm int}
=
-\lambda_i {\cal S}\bar D_L d_{Ri}+{\rm h.c.},
\qquad
{\cal S}=S_R\;{\rm or}\;\Phi,
\qquad i=s,b,
\label{eq:Linteraction}
\end{equation}
Equivalently, using $\bar D_L=\bar D P_R$, this interaction can be written as
\begin{equation}
\mathcal L_{\rm int}
=
-\lambda_i{\cal S}\bar D P_Rd_i
-\lambda_i^\ast {\cal S}^\dagger \bar d_iP_LD .
\end{equation}
For the real benchmark, ${\cal S}^\dagger={\cal S}=S_R$; for the complex benchmark, ${\cal S}^\dagger=\Phi^\dagger$ is distinct from $\Phi$.
In a general renormalizable theory, gauge invariance can allow other interactions as well. For example, the $D$ can mix with the SM down-type quarks through Higgs ($H$) terms such as $\bar q_L H D_R$. The $S$ can also have Higgs-portal interactions, for example $S H^\dagger H$ and $S^2H^\dagger H$. These terms are not part of the benchmark studied here. A simple way to obtain this restricted benchmark is to impose a $Z_2$ parity. Under this symmetry, the new fields $S$ and $D$ are odd, while all SM fields are even. In this benchmark we assume that possible direct mixing terms between the vector-like quark and the SM down-type quarks are absent. The interaction $S\bar D_L d_{Ri}$ is then
allowed, whereas terms such as $\bar q_L H D_R$ and $\bar D_L d_R$ are not. We also take $\langle S\rangle=0$, so that the same interaction $S\bar D_L d_{Ri}$ does not generate mass mixing after symmetry breaking. The Higgs-portal term $S^2H^\dagger H$ may be present, but it only affects the scalar sector through the physical value of $m_S$ and does not enter the leading $S$--$D$ dipole matching, which will be discussed later.

\begin{figure}[H]
\centering

\tikzset{
  fermion/.style={
    line width=0.8pt,
    postaction={decorate},
    decoration={
      markings,
      mark=at position 0.58 with {\arrow{>}}
    }
  },
  scalar/.style={line width=0.8pt,dashed},
  photon/.style={
    line width=0.8pt,
    decorate,
    decoration={snake,amplitude=0.8mm,segment length=3mm}
  },
  gluon/.style={
    line width=0.8pt,
    decorate,
    decoration={coil,amplitude=0.9mm,segment length=2.5mm}
  },
  vtx/.style={circle,fill=black,inner sep=1.4pt},
  vf/.style={font=\scriptsize},
  lab/.style={font=\small}
}

\resizebox{0.55\textwidth}{!}{%
\begin{tabular}{cc}

%==================================================
% (a) X \bar D d_i
%==================================================
\begin{tikzpicture}[baseline=(current bounding box.center)]
  \coordinate (v) at (0,0);
  \coordinate (S) at (0,1.2);
  \coordinate (Db) at (-1.3,-0.95);
  \coordinate (di) at (1.3,-0.95);

  \draw[scalar] (S) -- (v);
  \draw[fermion] (Db) -- (v);
  \draw[fermion] (v) -- (di);

  \node[vtx] at (v) {};
  \node[above] at (S) {${\cal S}$};
  \node[left]  at (Db) {$\bar D$};
  \node[right] at (di) {$d_i$};

  \node[vf,above right] at (v) {$-i\lambda_i P_R$};
  \node[lab] at (0,-1.55) {(a)};
\end{tikzpicture}
&
%==================================================
% (b) X^\dagger \bar d_i D
%==================================================
\begin{tikzpicture}[baseline=(current bounding box.center)]
  \coordinate (v) at (0,0);
  \coordinate (S) at (0,1.2);
  \coordinate (db) at (-1.3,-0.95);
  \coordinate (D) at (1.3,-0.95);

  \draw[scalar] (S) -- (v);
  \draw[fermion] (db) -- (v);
  \draw[fermion] (v) -- (D);

  \node[vtx] at (v) {};
  \node[above] at (S) {${\cal S}^\dagger$};
  \node[left]  at (db) {$\bar d_i$};
  \node[right] at (D) {$D$};

  \node[vf,above right] at (v) {$-i\lambda_i^\ast P_L$};
  \node[lab] at (0,-1.55) {(b)};
\end{tikzpicture}

\\[1.0cm]

%==================================================
% (c) D \bar D \gamma
%==================================================
\begin{tikzpicture}[baseline=(current bounding box.center)]
  \coordinate (v) at (0,0);
  \coordinate (Db) at (-1.3,-0.95);
  \coordinate (D) at (-1.3,0.95);
  \coordinate (ga) at (1.45,0);

  \draw[fermion] (Db) -- (v);
  \draw[fermion] (v) -- (D);
  \draw[photon] (v) -- (ga);

  \node[vtx] at (v) {};
  \node[left]  at (Db) {$\bar D$};
  \node[left]  at (D) {$D$};
  \node[right] at (ga) {$\gamma$};

  \node[vf,above right] at (v) {$ieQ_D\gamma^\mu$};
  \node[lab] at (0,-1.55) {(c)};
\end{tikzpicture}
&
%==================================================
% (d) D \bar D g^a
%==================================================
\begin{tikzpicture}[baseline=(current bounding box.center)]
  \coordinate (v) at (0,0);
  \coordinate (Db) at (-1.3,-0.95);
  \coordinate (D) at (-1.3,0.95);
  \coordinate (gl) at (1.45,0);

  \draw[fermion] (Db) -- (v);
  \draw[fermion] (v) -- (D);
  \draw[gluon] (v) -- (gl);

  \node[vtx] at (v) {};
  \node[left]  at (Db) {$\bar D$};
  \node[left]  at (D) {$D$};
  \node[right] at (gl) {$g^a$};

  \node[vf,above right] at (v) {$ig_s T^a\gamma^\mu$};
  \node[lab] at (0,-1.55) {(d)};
\end{tikzpicture}

\end{tabular}
}

\caption{Feynman rule vertices for the benchmark model. In panels (a) and (b), ${\cal S}$ denotes the neutral singlet scalar: ${\cal S}=S_R$ for the real-scalar benchmark and ${\cal S}=\Phi$ for the complex-scalar benchmark. Panels (c) and (d) show the electromagnetic and chromodynamic couplings of the vector-like quark $D$.}
\label{feynman_vertices_benchmark}
\end{figure}
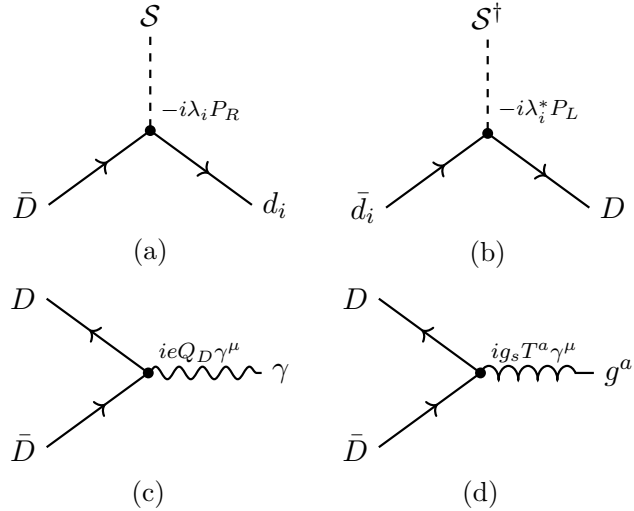

The interaction in Eq.~\eqref{eq:Linteraction} is gauge invariant, and its relevant gauge quantum number is mentioned in Table~\ref{gauge_quantum_numbers}.
\begin{table}[H]
\centering
\caption{Fields transformation under the SM gauge group.}
\label{gauge_quantum_numbers}
\renewcommand{\arraystretch}{1.2}
\setlength{\tabcolsep}{10pt}
\rowcolors{2}{blue!5}{white}
\begin{tabular}{ccccc}
\toprule
\rowcolor{blue!15}
\textbf{Field} & \textbf{$SU(3)_c$} & \textbf{$SU(2)_L$} & \textbf{$U(1)_Y$} & \textbf{Representation} \\
\midrule
$S_R=S_R^\dagger$ or $\Phi$ & $1$ & $1$ & $0$ & $(1,1,0)$ \\
$D_L$                       & $3$ & $1$ & $-1/3$ & $(3,1,-1/3)$ \\
$D_R$                       & $3$ & $1$ & $-1/3$ & $(3,1,-1/3)$ \\
$\bar D_L$                  & $\bar 3$ & $1$ & $+1/3$ & $(\bar 3,1,+1/3)$ \\
$d_{Ri}$                    & $3$ & $1$ & $-1/3$ & $(3,1,-1/3)$ \\
\bottomrule
\end{tabular}
\end{table}
For the color part,
\begin{equation*}
1\otimes \bar 3\otimes 3 \supset 1,
\end{equation*}
for the weak-isospin part,
\begin{equation*}
1\otimes 1\otimes 1 =1,
\end{equation*}
and for hypercharge,
\begin{equation*}
0+\frac13-\frac13=0.
\end{equation*}
Therefore,
\begin{equation*}
(1,1,0)\otimes(\bar 3,1,+1/3)\otimes(3,1,-1/3)
\supset
(1,1,0),
\end{equation*}
so the operator $S_R\bar D_L d_{Ri}$ in the real-scalar benchmark and the operator
$\Phi\bar D_L d_{Ri}$ in the complex-scalar benchmark are both SM gauge singlets.
In the complex-scalar benchmark, the hermitian-conjugate interaction is $-\lambda_i^{\ast}\Phi^\dagger \bar d_{Ri}D_L$,
which is also gauge invariant since
\begin{equation*}
(1,1,0)\otimes(\bar 3,1,+1/3)\otimes(3,1,-1/3)
\supset
(1,1,0),
\end{equation*}
and is therefore gauge invariant.
The FCNC transition is controlled by the product $\lambda_s^\ast\lambda_b$. The corresponding Feynman rules are given in Figure~\ref{feynman_vertices_benchmark}.

Because ${\cal S}$ is both electrically neutral and a color singlet, the photon and the gluon in the one-loop $b\to s\gamma$ and $b\to sg$ amplitudes can only be emitted from the internal vector-like quark $D$ line can be seen in Figure~\ref{loopdiagrams}. This observation will be important for understanding the relation between the matching functions for $C_{7\gamma}$ and $C_{8G}$.

\section{Effective Hamiltonian for {$b\to s\gamma$}}
\label{sec:hamiltonian}

The low-energy effective Hamiltonian relevant for $b\to s\gamma$ can be written as~\cite{Buchalla:1995vs,Buras:1998raa,Grinstein:1990tj,Misiak:1992bc}
\begin{equation}
\mathcal H_{\rm eff}
=
-\frac{4G_F}{\sqrt2}V_{tb}V_{ts}^\ast
\left[
C_{7\gamma}O_{7\gamma}
+
C_{8G}O_{8G}
+\cdots
\right].
\label{Heff}
\end{equation}
Using the standard operator basis for radiative decays~\cite{Grinstein:1990tj,Misiak:1992bc,Chetyrkin:1996vx}, the electromagnetic and chromomagnetic dipole operators are defined as
\begin{align}
O_{7\gamma}
=&
\frac{e}{16\pi^2}m_b
\left(
\bar s\sigma^{\mu\nu}P_R b
\right)F_{\mu\nu},
\label{eq:O7}\\
O_{8G}
=&
\frac{g_s}{16\pi^2}m_b
\left(
\bar s\sigma^{\mu\nu}T^aP_R b
\right)G^a_{\mu\nu}.
\label{eq:O8}
\end{align}
Here $F_{\mu\nu}$ and $G^a_{\mu\nu}$ are the electromagnetic and gluon field-strength tensors, respectively, and $P_R=(1+\gamma_5)/2$. The operator $O_{7\gamma}$ directly mediates the radiative transition $b\to s\gamma$, while $O_{8G}$ contributes indirectly through operator mixing under QCD running. Therefore, a consistent NP calculation of $B\to X_s\gamma$ requires both $C_{7\gamma}^{\rm NP}$ and $C_{8G}^{\rm NP}$.

The same convention is used below when we compare with the standard reference value~\cite{Buras:1998raa,Misiak:1992bc,Misiak:2020vlo}
\begin{equation}
C_{7\gamma}^{\rm SM,eff}(\mu_b)\simeq -0.30.
\label{eq:C7SMref}
\end{equation}
Once this convention and the full theory Feynman rules are defined, the relative interference terms between the SM and NP amplitudes are measurable. We use a sign parameter $\xi=\pm1$ only to keep track of the overall sign of the loop amplitude. It is not an additional model parameter. The quantities that are independent of this sign convention are the loop function, the magnitude of the overall normalization, and the ratio $C_{7\gamma}^{\rm NP}/C_{8G}^{\rm NP}=Q_D$.

\section{One-loop Matching Computation}
\label{sec:matching}

\begin{figure}[H]
\centering
\begin{tikzpicture}[scale=1.0, every node/.style={font=\small}]
% First diagram: photon
\node at (0,-1) {$b\to s\gamma$};
\coordinate (b1) at (-2.4,0);
\coordinate (v1) at (-1.0,0);
\coordinate (v2) at (1.0,0);
\coordinate (s1) at (2.4,0);
\coordinate (top1) at (0,1.45);
\coordinate (gam1) at (0,2.3);

\draw[-{Latex[length=2mm]}, thick] (b1) -- (v1) node[midway,below] {$b$};
\draw[-{Latex[length=2mm]}, thick] (v2) -- (s1) node[midway,below] {$s$};
\draw[thick] (v1) -- (top1) node[midway,left] {$D$};
\draw[thick] (top1) -- (v2) node[midway,right] {$D$};
\draw[dashed, thick] (v1) -- (v2) node[midway,below] {${\cal S}$};
\draw[decorate, decoration={snake, amplitude=1.5mm, segment length=4mm}, thick] (top1) -- (gam1) node[above] {$\gamma$};
\filldraw (v1) circle (1.5pt);
\filldraw (v2) circle (1.5pt);
\filldraw (top1) circle (1.5pt);

% Second diagram: gluon
\begin{scope}[xshift=6.3cm]
\node at (0,-1) {$b\to sg$};
\coordinate (b2) at (-2.4,0);
\coordinate (u1) at (-1.0,0);
\coordinate (u2) at (1.0,0);
\coordinate (s2) at (2.4,0);
\coordinate (top2) at (0,1.45);
\coordinate (g2) at (0,2.3);

\draw[-{Latex[length=2mm]}, thick] (b2) -- (u1) node[midway,below] {$b$};
\draw[-{Latex[length=2mm]}, thick] (u2) -- (s2) node[midway,below] {$s$};
\draw[thick] (u1) -- (top2) node[midway,left] {$D$};
\draw[thick] (top2) -- (u2) node[midway,right] {$D$};
\draw[dashed, thick] (u1) -- (u2) node[midway,below] {${\cal S}$};
\draw[decorate, decoration={coil, amplitude=1.4mm, segment length=3.5mm}, thick] (top2) -- (g2) node[above] {$g$};
\filldraw (u1) circle (1.5pt);
\filldraw (u2) circle (1.5pt);
\filldraw (top2) circle (1.5pt);
\end{scope}
\end{tikzpicture}
\caption{Feynman one-loop diagrams for the electromagnetic and chromomagnetic dipole operators in the minimal singlet-scalar/vector-like-quark model. The dashed line denotes the singlet scalar ${\cal S}$, and the solid internal line denotes the vector-like quark $D$.}
\label{loopdiagrams}
\end{figure}
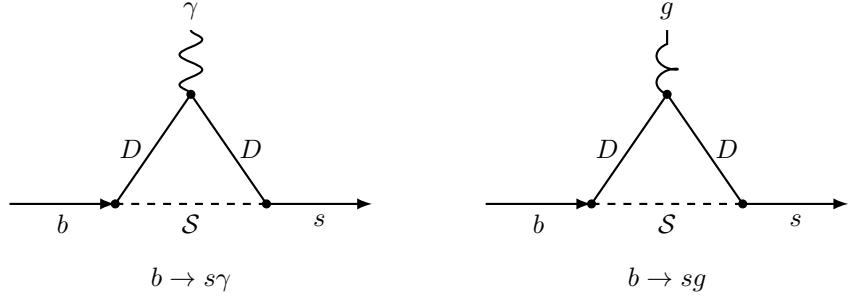

At the matching scale $\mu_D\sim M_D$, the new scalar ${\cal S}$, where ${\cal S}=S_R,\Phi$ and vector-like quark $D$ are integrated out. The relevant one-loop diagrams contain $D$ and $S$ in the loop, while the photon or gluon is emitted from the internal $D$ line. The two relevant one-loop Feynman diagrams are shown in Figure~\ref{loopdiagrams}. As the scalar $S$ is electrically neutral and color singlet, the photon and gluon are emitted only from the internal $D$ line. This is the reason why the electromagnetic and chromomagnetic matching functions are closely related to our minimal model.

This topology is a special case of the generic heavy scalar-fermion matching structures considered in the literature for $b\to s\gamma$ and $b\to sg$ transitions~\cite{Bobeth:1999ww}. In the present model, however, the scalar has $Q_S=0$ and is a color singlet. Therefore the terms that would arise from photon or gluon emission from the scalar line are absent, and only the vector-like-quark emission contribution remains. This is the origin of the simple relation between $C_{7\gamma}^{\rm NP}$ and $C_{8G}^{\rm NP}$ found below.

It is useful to explain how our notation is related to the generic notation used in Ref.~\cite{Bobeth:1999ww}. That work considers a heavy Dirac fermion $\psi$ and a heavy scalar $\phi$ with generic chiral couplings to the external $s$ and $b$ quarks, and matches them onto the same dipole operators used here. In that notation, contributions proportional to the product of same chirality couplings are conventionally denoted by an $R_1$ structure, while terms proportional to opposite chirality couplings are denoted by an $R_2$ structure and carry an internal chirality enhancement proportional to $m_\psi/m_b$. The minimal interaction in Eq.~\eqref{eq:Linteraction} contains only one chiral structure, so the $R_2$-type chirality-enhanced contribution is absent. Moreover, since our scalar is neutral and color singlet, the generic scalar-emission terms proportional to $Q_\phi$ or to the scalar color generator vanish. Thus, our result corresponds to the neutral, color-singlet scalar that limits the generic heavy scalar/fermion matching calculation.
We define the mass ratio
\begin{equation}
x=\frac{m_S^2}{M_D^2}.
\end{equation}
Using the generic scalar and fermion dipole matching formulae of Refs.~\cite{Bobeth:1999ww,Arnan:2019uhr}, and taking the neutral color-singlet scalar limit, the NP contribution to the WCs can be written as
\begin{align}
C_{7\gamma}^{\rm NP}(\mu_D)
=&
\xi\,
\frac{\lambda_s^\ast\lambda_b}
{4\sqrt2\,G_FV_{tb}V_{ts}^\ast M_D^2}
Q_DF_7(x),
\label{eq:C7match}\\
C_{8G}^{\rm NP}(\mu_D)
=&
\xi\,
\frac{\lambda_s^\ast\lambda_b}
{4\sqrt2\,G_FV_{tb}V_{ts}^\ast M_D^2}
F_8(x).
\label{eq:C8match}
\end{align}
The detailed computations for these NP WCs are provided in Appendix~\ref{sec:appendixB}. 
For the minimal model considered here, the scalar $S$ is electrically neutral and a color singlet. Therefore, in the notation of the generic scalar fermion matching formulae~\cite{Bobeth:1999ww,Arnan:2019uhr}, all terms corresponding to photon or gluon emission from the scalar line vanish. Only the gauge boson emission from the internal $D$
line remains. In this case the loop functions are equal,
\begin{equation}
F_7(x)=F_8(x)\equiv F(x),
\end{equation}
with
\begin{equation}
F(x)=
\frac{
1-6x+3x^2+2x^3-6x^2\ln x
}
{12(1-x)^4}.
\label{eq:loopfunction}
\end{equation}
This function has the finite limits
\begin{equation}
F(1)=\frac{1}{24},
\qquad
F(0)=\frac{1}{12}.
\end{equation}
For completeness, we briefly show how the loop function in Eq.~\eqref{eq:loopfunction} arises. The photon amplitude contains the generic loop integral
\begin{equation}
i\mathcal M_\gamma^\mu
\propto
eQ_D\,\lambda_s^\ast\lambda_b
\int\frac{d^d k}{(2\pi)^d}
\frac{
\bar u_s(p')
P_L(\slashed k+\slashed p'+M_D)\gamma^\mu
(\slashed k+\slashed p+M_D)P_Ru_b(p)
}
{\left[k^2-m_S^2\right]
\left[(k+p')^2-M_D^2\right]
\left[(k+p)^2-M_D^2\right]} .
\label{eq:fullamp_schematic}
\end{equation}
Here, the external momenta are $b(p)\to s(p^\prime)+\gamma(q)$ with $ q=p-p^\prime.$
The chromomagnetic amplitude has the same loop structure, with the replacement
$eQ_D\to g_sT^a$. We keep only the dipole part $\propto \bar u_s(p')\,i\sigma^{\mu\nu}q_\nu P_R\,u_b(p)$, and use the on-shell equations of motion, neglecting $m_s$ relative to $m_b$. The chirality flip comes from the external bottom-quark mass. There is no independent opposite chirality coupling in Eq.~\eqref{eq:Linteraction} therefore no internal chirality flip proportional to $M_D$ occurs. After combining the denominators with Feynman parameters, shifting the loop momentum, and keeping only the terms linear in the external photon momentum gives the form
\begin{align}
F(x)
&=
\frac{1}{2}\int_0^1 dz\,
\frac{z^2(1-z)}{x+(1-x)z}
\nonumber\\
&=
\frac{
1-6x+3x^2+2x^3-6x^2\ln x
}
{12(1-x)^4},
\end{align}
which is the expression quoted in Eq.~\eqref{eq:loopfunction}. 
The detailed computations for the loop function are given in Appendix~\ref{sec:appendixA}.
The relation between $C_{7\gamma}^{\rm NP}$ and $C_{8G}^{\rm NP}$ at the matching scale is then
\begin{equation}
C_{7\gamma}^{\rm NP}(\mu_D)
=
Q_D\,C_{8G}^{\rm NP}(\mu_D),
\end{equation}
up to the common normalization in Eqs.~\eqref{eq:C7match} and \eqref{eq:C8match}. Since $Q_D=-1/3$, the WCs $C_{7\gamma}^{\rm NP}(\mu_D)$ is reduced by the electric charge of the $D$ quark relative to the $C_{8G}^{\rm NP}(\mu_D)$ coefficient. This ratio is independent of the overall sign dependent on the convention.

\subsection{Leading-logarithmic running of the WCs}
\label{sec:running}

The WCs in Eqs.~\eqref{eq:C7match} and \eqref{eq:C8match} are defined at the high scale $\mu_D\sim M_D$. The heavy fields $D$ and $S$ are integrated out, and their SD effects are encoded in the WCs. To connect them with the physical decay $B\to X_s\gamma$, they must be evolved to the low scale $\mu_b\sim m_b$. This process is performed via standard computations using renormalization group equation (RGE). At leading-logarithmic order, following the standard $B\to X_s\gamma$ evolution are given in these Refs~\cite{Grinstein:1990tj,Misiak:1992bc,Chetyrkin:1996vx,Buchalla:1995vs,Buras:1998raa},
\begin{equation}
\mu\frac{d}{d\mu}C_i^{\rm eff}(\mu)
=
C_j^{\rm eff}(\mu)\gamma_{ji}^{\rm eff}(\mu).
\end{equation}
At leading order, the anomalous-dimension matrix is expanded as,
\begin{equation}
\hat\gamma^{\rm eff}(\mu)
=
\frac{\alpha_s(\mu)}{4\pi}\hat\gamma^{(0)\rm eff}+\cdots.
\end{equation}
Solving this leading-order RGE gives the standard expression for the electromagnetic dipole coefficient at $\mu_b$. In Eq.~(20) of Ref.~\cite{Chetyrkin:1996vx}, the leading-order solution is written as
\begin{equation}
C_7^{(0)\rm eff}(\mu_b)
=
\eta^{16/23}C_7^{(0)}(M_W)
+
\frac{8}{3}\left(\eta^{14/23}-\eta^{16/23}\right)C_8^{(0)}(M_W)
+
\sum_{i=1}^{8}h_i\eta^{a_i},
\label{eq:CMM20}
\end{equation}
where $\eta=\frac{\alpha_s(M_W)}{\alpha_s(\mu_b)}$.
The last term in Eq.~\eqref{eq:CMM20} comes from the SM four-quark operators. For the NP part considered here, we only consider the WCs $C_{7\gamma}^{\rm NP}(\mu_D)$ and $C_{8G}^{\rm NP}(\mu_D)$. Therefore, the four-quark contribution is not part of the NP only estimate. Replacing $M_W\to \mu_D$, the NP contribution evolves as
\begin{equation}
C_{7\gamma}^{\rm NP,eff}(\mu_b)
=
\eta^{16/23}C_{7\gamma}^{\rm NP}(\mu_D)
+
\frac{8}{3}\left(\eta^{14/23}-\eta^{16/23}\right)C_{8G}^{\rm NP}(\mu_D),
\label{eq:C7running}
\end{equation}
with
\begin{equation*}
\eta=\frac{\alpha_s(\mu_D)}{\alpha_s(\mu_b)}.
\end{equation*}
The same logic of matching at the high scale and evolving to the low scale is also described in Ref.~\cite{Misiak:1992bc}.
The second term in Eq.~\eqref{eq:C7running} shows explicitly that the chromomagnetic coefficient $C_{8G}$ contributes to the radiative decay through QCD mixing into $C_{7\gamma}$. Combining Eqs.~\eqref{eq:C7match}, \eqref{eq:C8match}, and \eqref{eq:C7running}, we obtain
\begin{equation}
C_{7\gamma}^{\rm NP,eff}(\mu_b)
=
\xi\,
\frac{\lambda_s^\ast\lambda_b}
{4\sqrt2\,G_FV_{tb}V_{ts}^\ast M_D^2}
F(x)
\left[
\eta^{16/23}Q_D
+
\frac{8}{3}
\left(
\eta^{14/23}-\eta^{16/23}
\right)
\right].
\label{eq:C7combined}
\end{equation}

\section{$B_s-\bar B_s$ Mixing}\label{sec:mixing}
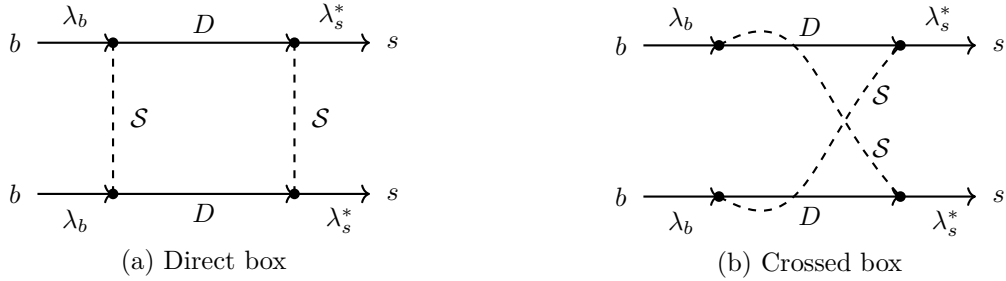
\begin{figure}[H]
\centering

\begin{minipage}{0.44\textwidth}
\centering
\begin{tikzpicture}[scale=1.0, every node/.style={font=\small}]

% External labels
\node at (-0.5,2) {$b$};
\node at (-0.5,0) {$b$};
\node at (4.5,2) {$s$};
\node at (4.5,0) {$s$};

% Internal vertices
\fill (0.8,2) circle (2pt);
\fill (0.8,0) circle (2pt);
\fill (3.2,2) circle (2pt);
\fill (3.2,0) circle (2pt);

% External fermion lines with simple arrows
\draw[->, thick] (-0.2,2) -- (0.8,2);
\draw[->, thick] (-0.2,0) -- (0.8,0);
\draw[->, thick] (3.2,2) -- (4.2,2);
\draw[->, thick] (3.2,0) -- (4.2,0);

% Internal fermion lines
\draw[->, thick] (0.8,2) -- (3.2,2);
\draw[->, thick] (0.8,0) -- (3.2,0);

% Scalar lines
\draw[dashed, thick] (0.8,2) -- (0.8,0);
\draw[dashed, thick] (3.2,2) -- (3.2,0);

% Labels
\node at (2.0,2.25) {$D$};
\node at (2.0,-0.25) {$D$};
\node at (1.15,1.0) {${\cal S}$};
\node at (3.55,1.0) {${\cal S}$};

\node at (0.3,2.35) {$\lambda_b$};
\node at (0.3,-0.35) {$\lambda_b$};
\node at (3.7,2.35) {$\lambda_s^\ast$};
\node at (3.8,-0.35) {$\lambda_s^\ast$};

\node at (2.0,-0.9) {(a) Direct box};

\end{tikzpicture}
\end{minipage}
\hspace{0.03\textwidth}
\begin{minipage}{0.44\textwidth}
\centering
\begin{tikzpicture}[scale=1.0, every node/.style={font=\small}]

% External labels
\node at (-0.5,2) {$b$};
\node at (-0.5,0) {$b$};
\node at (4.5,2) {$s$};
\node at (4.5,0) {$s$};

% Internal vertices
\fill (0.8,2) circle (2pt);
\fill (0.8,0) circle (2pt);
\fill (3.2,2) circle (2pt);
\fill (3.2,0) circle (2pt);

% External fermion lines
\draw[->, thick] (-0.2,2) -- (0.8,2);
\draw[->, thick] (-0.2,0) -- (0.8,0);
\draw[->, thick] (3.2,2) -- (4.2,2);
\draw[->, thick] (3.2,0) -- (4.2,0);

% Internal fermion lines
\draw[->, thick] (0.8,2) -- (3.2,2);
\draw[->, thick] (0.8,0) -- (3.2,0);

% Crossed scalar lines
\draw[dashed, thick] (0.8,2) .. controls (2.0,2.7) and (2.0,1.3) .. (3.2,0);
\draw[dashed, thick] (0.8,0) .. controls (2.0,-0.7) and (2.0,0.7) .. (3.2,2);

% Labels
\node at (2.0,2.25) {$D$};
\node at (2.0,-0.25) {$D$};
\node at (2.95,1.35) {${\cal S}$};
\node at (2.95,0.65) {${\cal S}$};

\node at (0.3,2.35) {$\lambda_b$};
\node at (0.3,-0.35) {$\lambda_b$};
\node at (3.7,2.35) {$\lambda_s^\ast$};
\node at (3.8,-0.35) {$\lambda_s^\ast$};

\node at (2.0,-0.9) {(b) Crossed box};

\end{tikzpicture}
\end{minipage}

\caption{One-loop box diagrams generating the $\Delta B=2$ transition $b\,b\to s\,s$ in the real and complex singlet-scalar with vector-like-quark benchmark.}
\label{fig:BsMixingBox}
\end{figure}
The coupling product $\lambda_s^\ast\lambda_b$ also enters $\Delta B=2$ box diagrams with two internal $D$ and two internal ${\cal S}$, where ${\cal S}=S_R,\Phi$ as shown in Figure~\ref{fig:BsMixingBox}. For the right-handed benchmark interaction, the leading contribution comes from the standard vector-right-right operator $O_{RR}^{bs}$ in the $\Delta B=2$ basis.~\cite{Buchalla:1995vs,Buras:1998raa,Lenz:2006hd,Arnan:2019uhr}:
\begin{equation}
O_{RR}^{bs}
=
(\bar s_\alpha\gamma_\mu P_R b_\alpha)
(\bar s_\beta\gamma^\mu P_R b_\beta),
\label{eq:ORRbs-main}
\end{equation}
the corresponding NP contribution to the $\Delta B=2$ Hamiltonian is written as
\begin{equation}
{\cal H}_{\rm eff}^{\Delta B=2,\,{\rm NP}}
= C_{RR}^{bs}(\mu)\,O_{RR}^{bs}(\mu)+{\rm h.c.}
\label{eq:Heff-db2-np-main}
\end{equation}
With the chiral projector convention $P_{R,L}=(1\pm\gamma_5)/2$, the matching WCs at the scale $\mu_D\simeq M_D$ is
\begin{equation}
C_{RR}^{bs,{\rm dir}}(\mu_D)
=
\frac{(\lambda_s^\ast\lambda_b)^2}{128\pi^2M_D^2}
G_{\rm box}(x),
\qquad
x=\frac{m_S^2}{M_D^2},
\label{eq:CRRbs-match-main}
\end{equation}
where
\begin{equation}
G_{\rm box}(x)=
\frac{1-x^2+2x\ln x}{(1-x)^3},
\qquad
G_{\rm box}(1)=\frac{1}{3}.
\label{eq:Gbox-main}
\end{equation}
The normalization in Eq.~\eqref{eq:CRRbs-match-main} is the standard scalar--fermion box normalization used in vector-like-fermion/singlet-scalar analyses~\cite{Grinstein:2018fgb,Arnan:2019uhr}; the same class of four-fermion operators and running factors is reviewed in Refs.~\cite{Buchalla:1995vs,Buras:1998raa,Lenz:2006hd}. In the notation commonly used for such box diagrams, one writes this as
\begin{equation}
C_{\Delta B=2}^{\rm dir}
=
\frac{(\lambda_s^\ast\lambda_b)^2}{128\pi^2m_S^2}\,F(r),
\qquad
r=\frac{M_D^2}{m_S^2}.
\label{eq:standard-box-reference-form}
\end{equation}
Since in the present paper we uses $x=m_S^2/M_D^2=1/r$, the same expression may be rewritten as
\begin{equation}
\frac{1}{m_S^2}F\!\left(\frac{1}{x}\right)
=
\frac{1}{M_D^2}G_{\rm box}(x),
\end{equation}
The $O_{RR}^{bs}$ operator has the same QCD anomalous dimension as the SM operator $O_{LL}^{bs}$. In leading logarithmic order,
\begin{equation}
C_{RR}^{bs}(\mu_b)
=
U_{\Delta B=2}(\mu_b,\mu_D)C_{RR}^{bs}(\mu_D),
\qquad
U_{\Delta B=2}
=
\left[\frac{\alpha_s(\mu_D)}{\alpha_s(\mu_b)}\right]^{6/23},
\label{eq:Udb2-main}
\end{equation}
as in the standard $\Delta B=2$ evolution of the current-current mixing operator~\cite{Buchalla:1995vs,Buras:1998raa,Lenz:2006hd}.
For comparison with the SM amplitude, we use the same chiral-projector normalization,
\begin{equation}
O_{LL}^{bs}
=
(\bar s_\alpha\gamma_\mu P_L b_\alpha)
(\bar s_\beta\gamma^\mu P_L b_\beta).
\end{equation}
In this convention the SM coefficient can be written as
\begin{equation}
C_{LL}^{\rm SM}(\mu_b)
=
\frac{G_F^2m_W^2}{4\pi^2}
(V_{tb}V_{ts}^\ast)^2\eta_B S_0(x_t),
\label{eq:CSMmix-main}
\end{equation}
where $S_0(x_t)$ is the Inami-Lim function~\cite{Inami:1980fz}, and $\eta_B$ is the SD QCD factor appearing in the SM $\Delta B=2$ amplitude~\cite{Buchalla:1995vs,Buras:1998raa,Lenz:2006hd}.

In the real-scalar benchmark, $S_R=S_R^\dagger$, we consider the contribution of both the direct box and crossed box Feynman graphs as shown in Figure~\ref{fig:BsMixingBox}. The crossed box contribution for real scalars is obtained from the scalar fermion expression by interchanging the scalar labels and replacing the relevant box loop function by $G\to -G$~\cite{Arnan:2019uhr}. In the present one real scalar, one vector like-quark benchmark, this interchange does not change the masses, couplings, or color factor. Therefore,
\begin{equation}
C_{RR}^{bs,{\rm cross}}(\mu_D)
=
-
\frac{(\lambda_s^{R\ast}\lambda_b^R)^2}{128\pi^2M_D^2}
G_{\rm box}(x),
\end{equation}
and the full leading result is
\begin{align}
C_{RR}^{bs,{\rm full}}(\mu_D)
&=
C_{RR}^{bs,{\rm dir}}(\mu_D)
+
C_{RR}^{bs,{\rm cross}}(\mu_D)
\nonumber\\
&=0.
\label{eq:real-scalar-cancellation-main}
\end{align}
Thus, the exact one real scalar benchmark does not have the leading dimension-six operator $O_{RR}^{bs}$ contribution from this ${\cal S}$-$D$ box. This cancellation applies to the single real scalar with identical direct and crossed Feynman loops. 
In the case of a complex scalar ${\cal S}=\Phi$, we have a conserved scalar current, $\langle\Phi\Phi^\dagger\rangle$, and therefore we are left with a contribution from the direct box Feynman graph. The details of the computation are provided in Appendix~\ref{sec:appendixC}.

\section{Numerical Analysis}
\label{sec:numerics}
In this section, we perform the numerical analysis.
We take $C_{7\gamma}^{\rm SM,eff}(\mu_b)$ as the reference SM coefficient and estimate the leading NP effect in the electromagnetic dipole approximation~\cite{Buras:1998raa,Misiak:2020vlo},
\begin{equation}
R_{X_s\gamma}
\equiv
\frac{\mathcal B(B\to X_s\gamma)}
{\mathcal B(B\to X_s\gamma)_{\rm SM}}
\simeq
\left|
1+
\frac{C_{7\gamma}^{\rm NP,eff}(\mu_b)}
{C_{7\gamma}^{\rm SM,eff}(\mu_b)}
\right|^2 .
\label{eq:Rratio}
\end{equation}
This expression is to give a clear first estimate. It includes the main dependence on the new WCs, but it should not be treated as a substitute for a full NNLO analysis of inclusive $B\to X_s\gamma$.

\subsection{Experimental Input and Constraints}
\label{subsec:expinput}

The current heavy-flavor average for the inclusive radiative branching fraction is~\cite{HeavyFlavorAveragingGroupHFLAV:2024ctg}
\begin{equation}
\mathcal B(B\to X_s\gamma)_{E_\gamma>1.6~{\rm GeV}}
=
(3.49\pm0.19)\times10^{-4},
\label{eq:hflav}
\end{equation}
while the NNLO SM prediction is~\cite{Misiak:2020vlo}
\begin{equation}
\mathcal B(B\to X_s\gamma)_{\rm SM}
=
(3.40\pm0.17)\times10^{-4}.
\label{eq:smbr}
\end{equation}
The experimental average and the SM prediction are therefore in good agreement. 
A useful way to show this agreement is to compare the inclusive measurement with the NNLO prediction~\cite{HeavyFlavorAveragingGroupHFLAV:2024ctg,Misiak:2020vlo}, is
\begin{equation}
R_{\rm exp/SM}
=
\frac{\mathcal B(B\to X_s\gamma)_{\rm exp}}
{\mathcal B(B\to X_s\gamma)_{\rm SM}}
=
1.03\pm0.08,
\label{eq:rexp_sm}
\end{equation}
where the uncertainty has been obtained by combining the experimental and SM uncertainties in quadrature.
There are two different numerical quantities that should be distinguished. First, Eq.~\eqref{eq:rexp_sm}, together with the simplified relation in Eq.~\eqref{eq:Rratio} and the reference value from Eq.~\eqref{eq:C7SMref}
suggests that a new contribution of order a few $10^{-2}$ in $C_{7\gamma}^{\rm NP,eff}$ would already be phenomenologically relevant. For orientation, we use
\begin{equation}
\left|C_{7\gamma}^{\rm NP,eff}(\mu_b)\right|
\lesssim 0.03
\label{eq:c7bound_data}
\end{equation}
To see the size of this estimate, define $r=C_{7\gamma}^{\rm NP,eff}/C_{7\gamma}^{\rm SM,eff}$.
For small NP,
\begin{align}
R_{X_s\gamma}=|1+r|^2
=1+2\,{\rm Re}(r)+|r|^2
\simeq 1+2\,{\rm Re}(r).
\end{align}
Thus a $\mathcal{O}(10\%)$ change in the branching ratio corresponds to an amplitude-level correction of order $\mathcal{O}(5\%)$. Allowing a somewhat conservative amplitude shift up to
$|r|\sim 0.1$ and using $|C_{7\gamma}^{\rm SM,eff}|\simeq 0.30$ gives $|C_{7\gamma}^{\rm NP,eff}|\sim 0.03$. We therefore use Eq.~\eqref{eq:c7bound_data} only as a data driven sensitivity estimate, not as a strict exclusion bound.

In addition, to show conservative benchmark plots, we use the loose radiative working criterion $\left|C_{7\gamma}^{\rm NP,eff}(\mu_b)\right|<0.05$. Therefore,
$<0.05$ is a plotting and orientation criterion, not a final experimental exclusion. Any coupling limit extracted using $\left|C_{7\gamma}^{\rm NP,eff}(\mu_b)\right|<0.05$ can be rescaled to a different assumed dipole sensitivity by replacing $0.05$ with the desired value. For example, using the indicative estimate in Eq.~\eqref{eq:c7bound_data} rescales these limits by$\frac{0.03}{0.05}\simeq0.6$.
A more refined analysis would require full perturbative treatment of the inclusive branching ratio, correlated experimental and theoretical uncertainties, and the full dependence on the WCs.

\begin{figure}[H]
\centering
\includegraphics[width=0.72\textwidth]{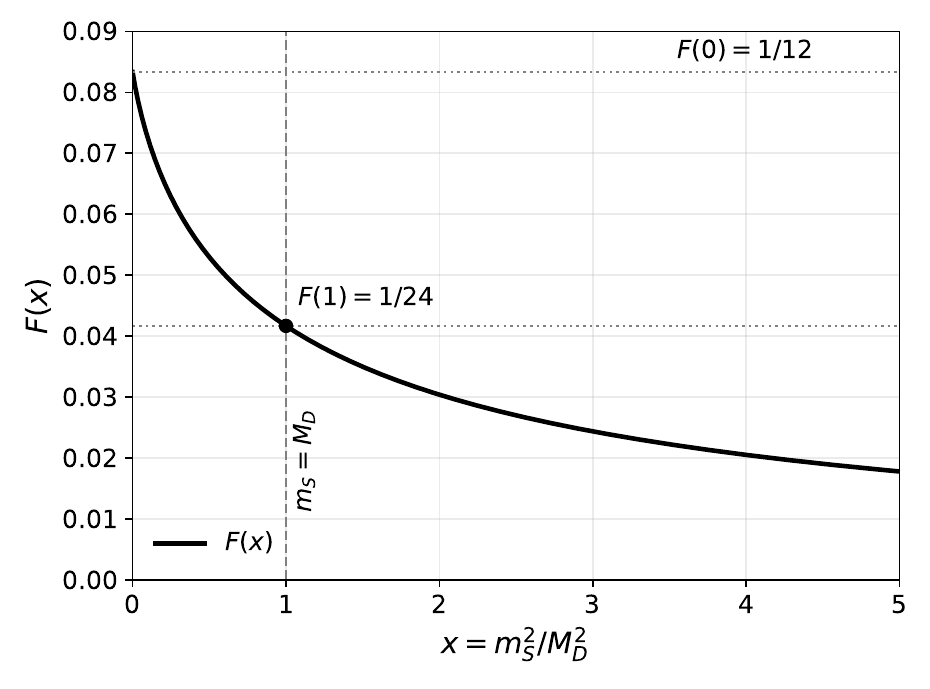}
\caption{Loop function $F(x)$ entering the matching contributions to $C_{7\gamma}^{\rm NP}$ and $C_{8G}^{\rm NP}$. The smooth decrease with $x=m_S^2/M_D^2$ shows the decoupling of a heavier singlet scalar $S$ relative to the vector-like down quark $D$.}
\label{fig:loopfunction}
\end{figure}

In addition, in our analysis we use the Fermi constant $G_F=1.1663787\times10^{-5}~{\rm GeV}^{-2}$~\cite{ParticleDataGroup:2024cfk} and the relevant CKM combination $|V_{tb}V_{ts}^\ast|=0.040$~\cite{ParticleDataGroup:2024cfk} 
together with Eq.~\eqref{eq:C7SMref}, in the same effective Hamiltonian convention as mention in Eq.~\eqref{Heff}. For the QCD running factor we use a one-loop expression with $n_f=5$, using the standard QCD normalization summarized by the Particle Data Group~\cite{ParticleDataGroup:2024cfk},
\begin{equation}
\alpha_s(\mu)=
\frac{\alpha_s(m_Z)}
{1+\dfrac{\beta_0\alpha_s(m_Z)}{2\pi}\ln(\mu/m_Z)},
\qquad
\beta_0=11-\frac{2n_f}{3},
\end{equation}
taking $\alpha_s(m_Z)=0.1180$ and $m_Z=91.1876~{\rm GeV}$~\cite{ParticleDataGroup:2024cfk}. This gives
\begin{equation}
\alpha_s(\mu_b=4.8~{\rm GeV})\simeq0.205,
\qquad
\alpha_s(1~{\rm TeV})\simeq0.0877,
\qquad
\eta\simeq0.428.
\end{equation}
Before the benchmark values, it is useful to show the behavior of the loop function when entering the matching WCs. In Figure~\ref{fig:loopfunction}, we present $F(x)$ as a function of
$x=m_S^2/M_D^2$. The $F(x)$ decreases smoothly as the singlet scalar $S$ mass increases relative to the vector-like down quark $D$ mass. In the limiting case of a light scalar, $x\to 0$, we finds $F(0)=1/12$, while for the equal-mass benchmark $m_S=M_D$, corresponding to $x=1$, the finite value is $F(1)=\frac{1}{24}\simeq 0.04167$.
This is the value used below for the benchmark point $M_D=m_S=1~{\rm TeV}$ and $|\lambda_s^\ast\lambda_b|=1$.
Using Eqs.~\eqref{eq:C7match} and \eqref{eq:C8match}, the magnitudes at the matching scale are then given by
\begin{equation}
\left|C_{7\gamma}^{\rm NP}(\mu_D)\right|\simeq 5.3\times10^{-3},
\qquad
\left|C_{8G}^{\rm NP}(\mu_D)\right|\simeq 1.58\times10^{-2}.
\label{eq:benchmark-high-scale-magnitudes}
\end{equation}
Putting these values into the leading logarithmic running to the low scale, Eq.~\eqref{eq:C7running} gives
\begin{equation}
\left|C_{7\gamma}^{\rm NP,eff}(\mu_b)\right|
\simeq 1.13\times10^{-3}.
\label{eq:benchmark-low-scale-magnitude}
\end{equation}
The NP WC Relative to the SM WC is given by
\begin{equation}
\epsilon_7
\equiv
\left|
\frac{C_{7\gamma}^{\rm NP,eff}(\mu_b)}
{C_{7\gamma}^{\rm SM,eff}(\mu_b)}
\right|
\simeq
3.8\times10^{-3}.
\label{eq:epsilon7benchmark}
\end{equation}
Thus the benchmark shift is below the half-percent level at the amplitude level. If the new contribution interferes linearly with the SM amplitude, Eq.~\eqref{eq:Rratio} gives a branching-ratio shift of order
\begin{equation}
\Delta R_{X_s\gamma}
\equiv R_{X_s\gamma}-1
\simeq
2\,\mathrm{Re}\!
\left(
\frac{C_{7\gamma}^{\rm NP,eff}}
{C_{7\gamma}^{\rm SM,eff}}
\right)
\sim \pm0.76\%
\label{eq:deltaRbenchmark}
\end{equation}
This shows why the minimal benchmark is still safely consistent with the present $B\to X_s\gamma$ experimental and SM agreement for TeV-scale masses and $\mathcal{O}(1)$ $\sim |\lambda_s^\ast\lambda_b|$.

\begin{figure}[H]
\centering
\includegraphics[width=0.72\textwidth]{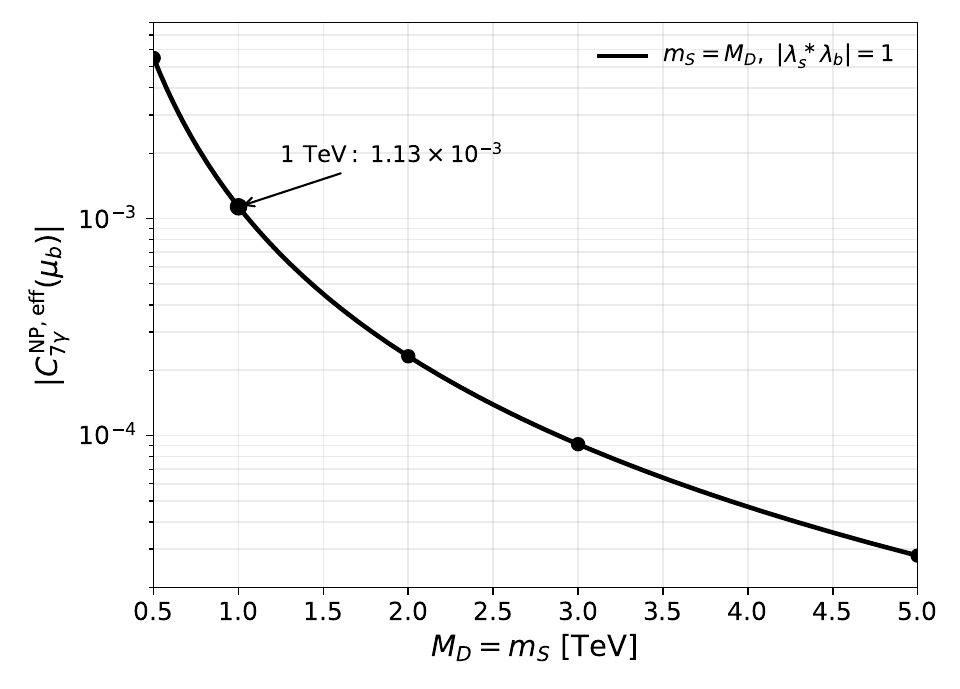}
\caption{Magnitude of the low scale NP contribution $\left|C_{7\gamma}^{\rm NP,eff}(\mu_b)\right|$ as a function of $M_D$.}
\label{fig:c7eff}
\end{figure}

The dependence of $\left|C_{7\gamma}^{\rm NP,eff}(\mu_b)\right|$ on $M_D$ is shown in Figure~\ref{fig:c7eff}. For fixed $m_S=M_D$ and $|\lambda_s^\ast\lambda_b|=1$, the contribution decreases rapidly with increasing $M_D$, showing the expected decoupling behavior proportional to $1/M_D^2$.

\begin{figure}[H]
\centering
\includegraphics[width=0.72\textwidth]{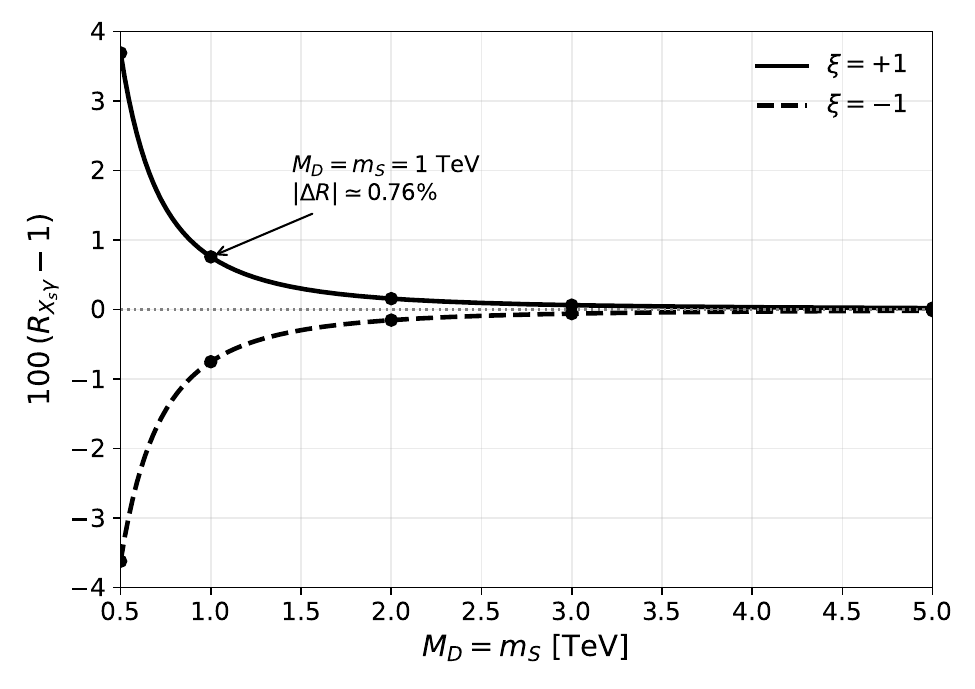}
\caption{Approximate percentage shift in the branching-ratio ratio,
$100\,(R_{X_s\gamma}-1)$, as a function of the mass
$M_D=m_S$, for $|\lambda_s^\ast\lambda_b|=1$.
}
\label{fig:rxsgamma}
\end{figure}

The corresponding effect on the approximate $\Delta R_{X_s\gamma}$ is shown in Figure~\ref{fig:rxsgamma}. In Figure~\ref{fig:rxsgamma} we observe the corresponding percentage shift in the $\Delta R_{X_s\gamma}$. For the benchmark with $M_D=m_S=1~{\rm TeV}$ and $|\lambda_s^\ast\lambda_b|=1$, the shift is only of order $\pm 0.76\%$. This is much smaller than the present experimental and SM uncertainty in $B\to X_s\gamma$, confirming that the minimal benchmark is radiatively safe at the TeV scale. The two curves correspond to the two choices of the sign parameter $\xi=\pm1$ used in the matching formulae. These curves should be read as an interference diagnostic rather than as two independent physical models: once a definite amplitude convention is fixed, only one curve corresponds to the Lagrangian in Eq.~\eqref{eq:Linteraction}. One sign gives a small enhancement relative to the SM value, while the opposite sign gives a small suppression. In both cases, the ratio approaches unity as $M_D$ increases.

\begin{figure}[H]
\centering
\includegraphics[width=0.60\textwidth]{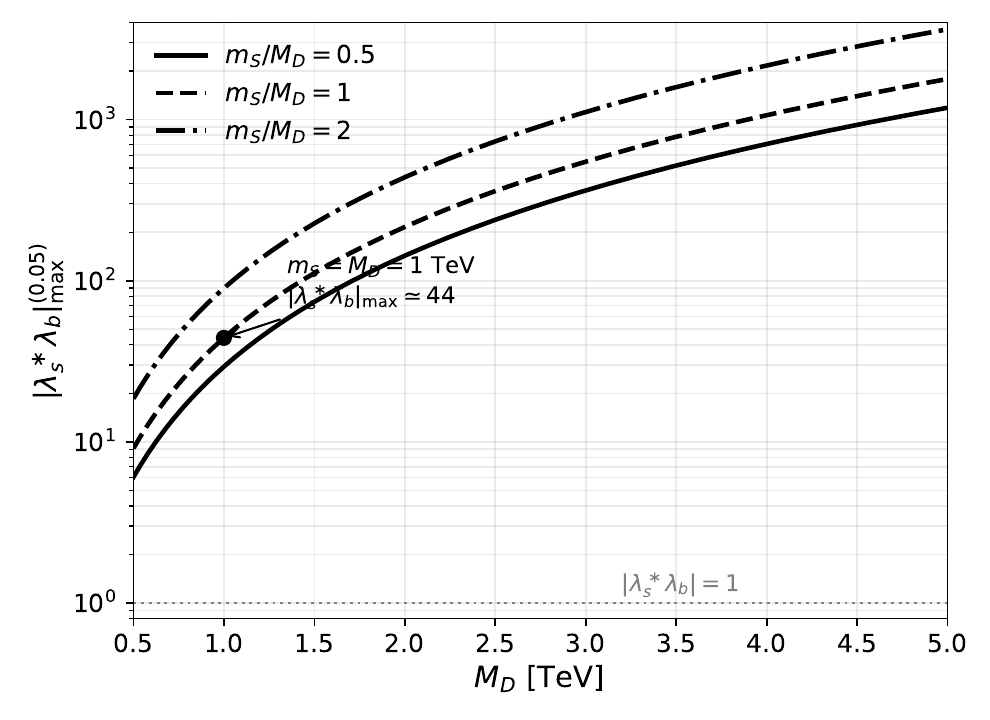}
\caption{Illustrative radiative-only upper limit on the coupling product
$|\lambda_s^\ast\lambda_b|$ as a function of $M_D$, obtained from the
working condition $|C_{7\gamma}^{\rm NP,eff}(\mu_b)|<0.05$. The three
curves correspond to different mass ratios $m_S/M_D$.}
\label{fig:lambdamax}
\end{figure}

Using the conservative working criterion $<0.05$, we obtain an illustrative radiative-only upper limit on the $|\lambda_s^\ast\lambda_b|$,
\begin{equation}
|\lambda_s^\ast\lambda_b|_{\rm max}^{(0.05)}
=
\frac{0.05}
{\left|C_{7\gamma}^{\rm NP,eff}(\mu_b)\right|_{|\lambda_s^\ast\lambda_b|=1}}.
\label{eq:lambdamax005}
\end{equation}
If instead the indicative data-driven sensitivity in Eq.~\eqref{eq:c7bound_data} is used, the corresponding estimate is
\begin{equation}
|\lambda_s^\ast\lambda_b|_{\rm max}^{(0.03)}
=
\frac{0.03}{0.05}
|\lambda_s^\ast\lambda_b|_{\rm max}^{(0.05)}
\simeq
0.6\,|\lambda_s^\ast\lambda_b|_{\rm max}^{(0.05)}.
\label{eq:lambdamax003}
\end{equation}
For the benchmark point $M_D=m_S=1~{\rm TeV}$, and put Eq.~\eqref{eq:benchmark-low-scale-magnitude} gives
\begin{equation}
|\lambda_s^\ast\lambda_b|_{\rm max}^{(0.05)}\simeq44,
\qquad
|\lambda_s^\ast\lambda_b|_{\rm max}^{(0.03)}\simeq27.
\label{eq:lambdamaxbenchmarkvalues}
\end{equation}
These large numbers should not be interpreted as realistic preferred couplings. Rather, they mean that $B\to X_s\gamma$ alone is not the limiting constraint on this minimal benchmark when the new particles are near the TeV scale.
In Figure~\ref{fig:lambdamax}, we observe that the radiative only upper limit on $|\lambda_s^\ast\lambda_b|$ obtained from the conservative condition $|C_{7\gamma}^{\rm NP,eff}(\mu_b)|<0.05$. Since $C_{7\gamma}^{\rm NP,eff}$ is linear in the coupling product, the limit is obtained by rescaling the unit-coupling result. The allowed value increases rapidly with $M_D$, which shows the decoupling of the heavy particles. For larger $m_S/M_D$, the $F(x)$ is smaller, so the radiative contribution is further suppressed and the corresponding radiative only limit becomes weaker.

\begin{figure}[H]
\centering
\includegraphics[width=0.60\textwidth]{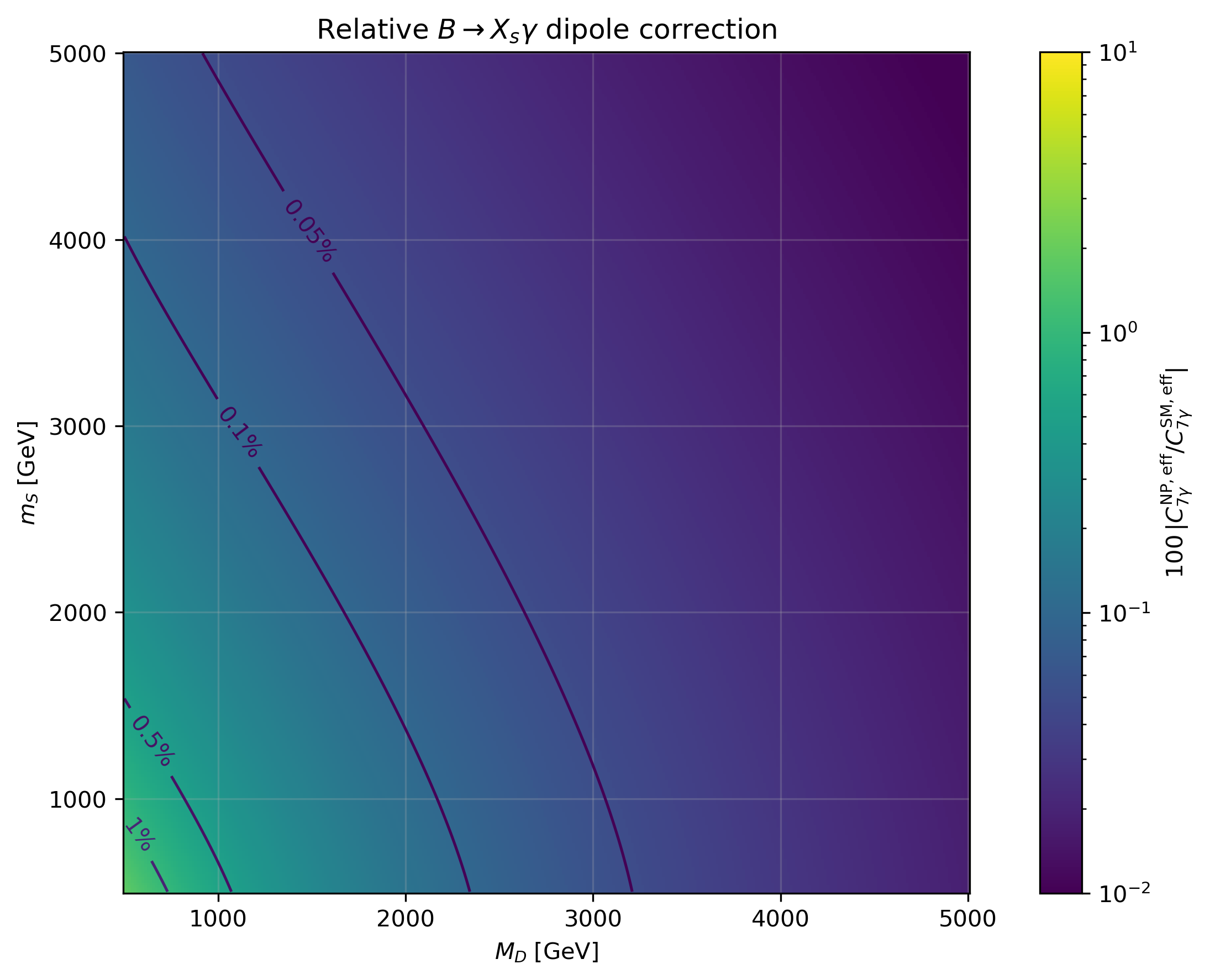}
\caption{2D scan of the relative NP contribution to SM,
$100\,|C_{7\gamma}^{\rm NP,eff}(\mu_b)/C_{7\gamma}^{\rm SM,eff}(\mu_b)|$, in the $(M_D,m_S)$ plane for $|\lambda_s^\ast\lambda_b|=1$.}
\label{fig:2drelative}
\end{figure}

A complementary two dimensional (2D) scan in the $(M_D,m_S)$ plane is shown in Figure~\ref{fig:2drelative}. Instead of showing only an allowed/excluded region, in this figure we see the relative size of the new WCs compared with the SM WCs,
\begin{equation}
100\times
\left|
\frac{C_{7\gamma}^{\rm NP,eff}(\mu_b)}
{C_{7\gamma}^{\rm SM,eff}(\mu_b)}
\right|.
\label{eq:relativepercent}
\end{equation}
This is more useful for the minimal benchmark because most of the TeV-scale parameter space satisfies the conservative radiative bound. The effect is largest when both $D$ and $S$ are relatively light, and it quickly becomes smaller as either mass increases. The contours indicate corrections of $1\%$, $5\%$, and $10\%$ relative to the SM dipole amplitude. As we see that the regions below the $1\%$ contour correspond to corrections smaller than the benchmark sensitivity of the present inclusive radiative observable.

This benchmark illustrates an important feature of the minimal model. The new contribution is suppressed by the loop factor and by $1/M_D^2$. Moreover, because the interaction in Eq.~\eqref{eq:Linteraction} contains only one chiral structure, there is no chirality enhancement proportional to $M_D/m_b$. Therefore, the correction to $C_{7\gamma}^{\rm NP,eff}(\mu_b)$ is small for TeV-scale vector-like quarks, even when the $|\lambda_s^\ast\lambda_b|$ is of $\mathcal{O}(1)$.

\subsection{$B_s-\bar B_s$ Mixing Constraint in the Complex-Scalar Benchmark}
\label{subsec:bs-mixing-role}

\begin{figure}[H]
\centering
\includegraphics[width=0.72\textwidth]{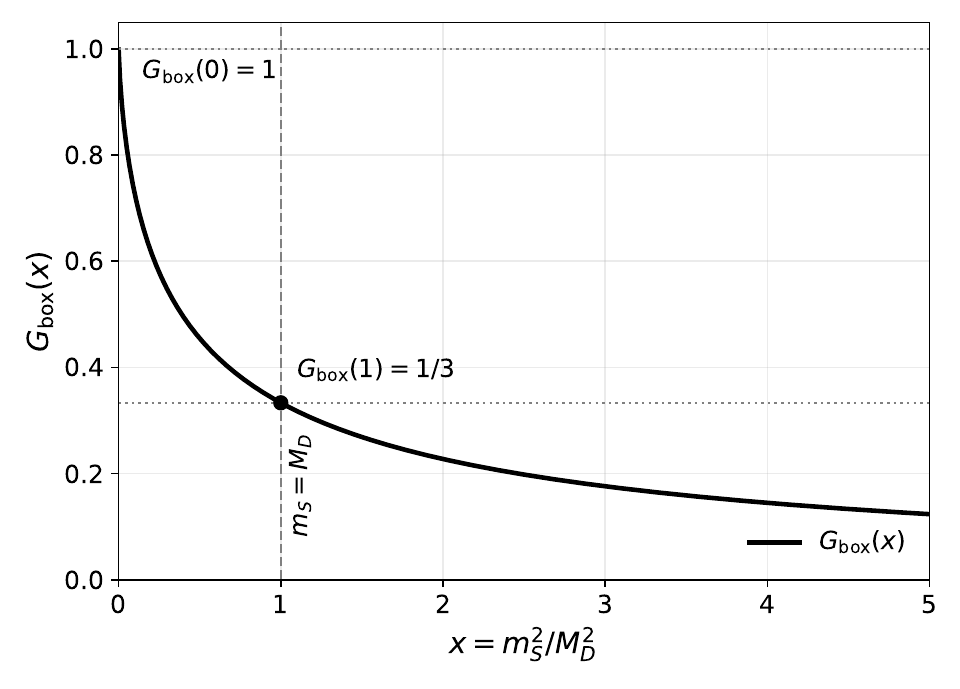}
\caption{$G_{\rm box}(x)$ entering the $B_s-\bar B_s$ mixing coefficient, with $x=m_S^2/M_D^2$. The finite value $G_{\rm box}(1)=1/3$ is given by the equal mass benchmark. The decrease at larger $x$ shows that a heavier ${\cal S}$ suppresses the $\Delta B=2$ box amplitude.}
\label{fig:gbox}
\end{figure}

The fact that $B\to X_s\gamma$ gives a weak constraint in our benchmarks does not mean that the model is fully safe from flavor constraints. The same $\lambda_s^\ast\lambda_b$ also appears in $\Delta B=2$ box diagram with two internal $D$ and two internal ${\cal S}=\Phi$. Therefore, we also take this into account for further analysis.
Figure~\ref{fig:gbox} shows that the box loop function $G_{\rm box}(x)$ entering the $\Delta B=2$ WCs for $B_s-\bar B_s$ mixing. The function is positive and decreases as $x$
increases, showing that a heavier scalar suppresses the box contribution. For the equal-mass benchmark $m_S=M_D$, one has $x=1$ and $G_{\rm box}(1)=1/3$.

Using $\mu_D=1~{\rm TeV}$ and $\mu_b=4.8~{\rm GeV}$ in Eq.~\eqref{eq:Udb2-main} gives $U_{\Delta B=2}\simeq0.80$.
Numerically, using $S_0(x_t)\simeq2.31$, $\eta_B\simeq0.55$, and $|V_{tb}V_{ts}^\ast|=0.040$, Eq.~\eqref{eq:CSMmix-main} gives
\begin{equation}
|C_{LL}^{\rm SM}(\mu_b)|\simeq4.5\times10^{-11}~{\rm GeV}^{-2}.
\label{eq:CSMmix-number}
\end{equation}
Where $S_0(x_t)$ is the Inami Lim box function~\cite{Inami:1980fz}, evaluated at $x_t=\bar m_t^2(m_t)/m_W^2$. Using $\bar m_t(m_t)\simeq 163~\mathrm{GeV}$ and $m_W\simeq 80.4~\mathrm{GeV}$, we obtain $S_0(x_t)\simeq 2.31$. The matrix elements of $O_{RR}^{bs}$ and $O_{LL}^{bs}$ are equal by parity, as used in standard $\Delta B=2$ analyses of left- and right-handed vector operators~\cite{Buras:1998raa,Lenz:2006hd,Arnan:2019uhr}. Therefore, the relative correction to the mixing amplitude can be estimated by the WCs ratio
\begin{equation}
h_s\equiv
\left|\frac{C_{RR}^{bs}(\mu_b)}{C_{LL}^{\rm SM}(\mu_b)}\right|.
\label{eq:hsdef}
\end{equation}
$h_s$ is the appropriate measure of the fractional NP correction. For complex $\lambda_s^\ast\lambda_b$, the same expression gives the magnitude of the correction, while the phase would also affect the $B_s$ mixing phase.

The experimental mass difference is very precisely measured, $\Delta M_s^{\rm exp}=17.765\pm0.006~{\rm ps}^{-1}$~\cite{HeavyFlavorAveragingGroupHFLAV:2024ctg,ParticleDataGroup:2024cfk}. The SM prediction is limited mainly by CKM and lattice inputs, as emphasized in modern $B_s$-mixing studies~\cite{Lenz:2006hd,DiLuzio:2019jyq}; for example Ref.~\cite{DiLuzio:2019jyq} quotes $\Delta M_s^{\rm SM}=18.4^{+0.7}_{-1.2}~{\rm ps}^{-1}$. We take 
\begin{equation}
h_s < \delta_{B_s},
\qquad
\delta_{B_s}=0.20,
\label{eq:deltabscriterion}
\end{equation}
we take this condition that should be interpreted as a conservative criterion,
not as a statistically defined exclusion limit. We also quote the rescaling criteria to $10\%$ and $5\%$ below.

\begin{figure}[H]
\centering
\includegraphics[width=0.72\textwidth]{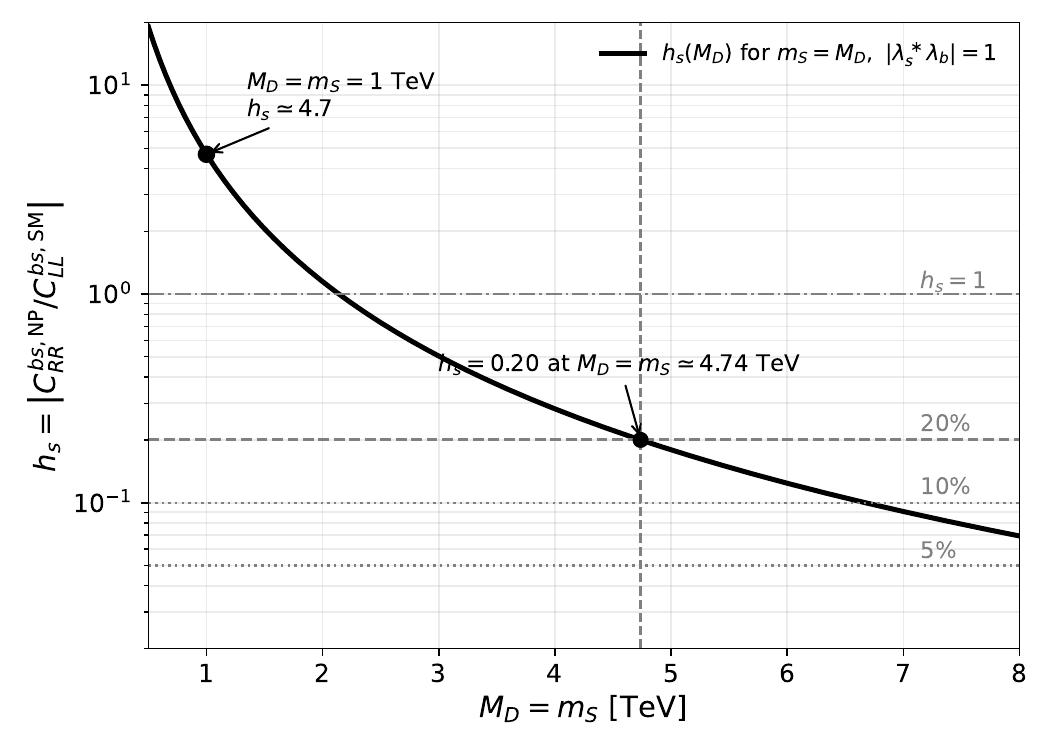}
\caption{Relative $B_s-\bar B_s$ mixing contribution $h_s=|C_{RR}^{bs,\rm NP}/C_{LL}^{bs,\rm SM}|$ as a function of $M_D$ for $m_S=M_D$ and $|\lambda_s^\ast\lambda_b|=1$.}
\label{fig:bsmixing-ratio}
\end{figure}

For the benchmark point $M_D=m_S=1~{\rm TeV}$ and $|\lambda_s^\ast\lambda_b|=1$, we find
\begin{equation}
h_s\simeq 4.7.
\label{eq:hs-lamone}
\end{equation}
The Figure~\ref{fig:bsmixing-ratio} show the relative NP contribution to
$B_s-\bar B_s$ mixing with equal mass line $m_S=M_D$ for $|\lambda_s^\ast\lambda_b|=1$. Since $x=m_S^2/M_D^2=1$ on this line, the $G(x)$ is fixed at $G_{\rm box}(1)=1/3$, and the decrease of the curve mainly follows the expected $1/M_D^2$ decoupling behavior. The benchmark point $M_D=m_S=1~{\rm TeV}$ gives $h_s\simeq4.7$, this means that the new contribution is several times larger than the SM mixing amplitude. For this reason, this point is not compatible with the $h_s<0.20$. For an $\mathcal{O}{(1)}\sim |\lambda_s^\ast\lambda_b|$, the equal-mass spectrum becomes compatible with this criterion only for masses of order several TeV.

\begin{figure}[H]

\centering
\includegraphics[width=0.72\textwidth]{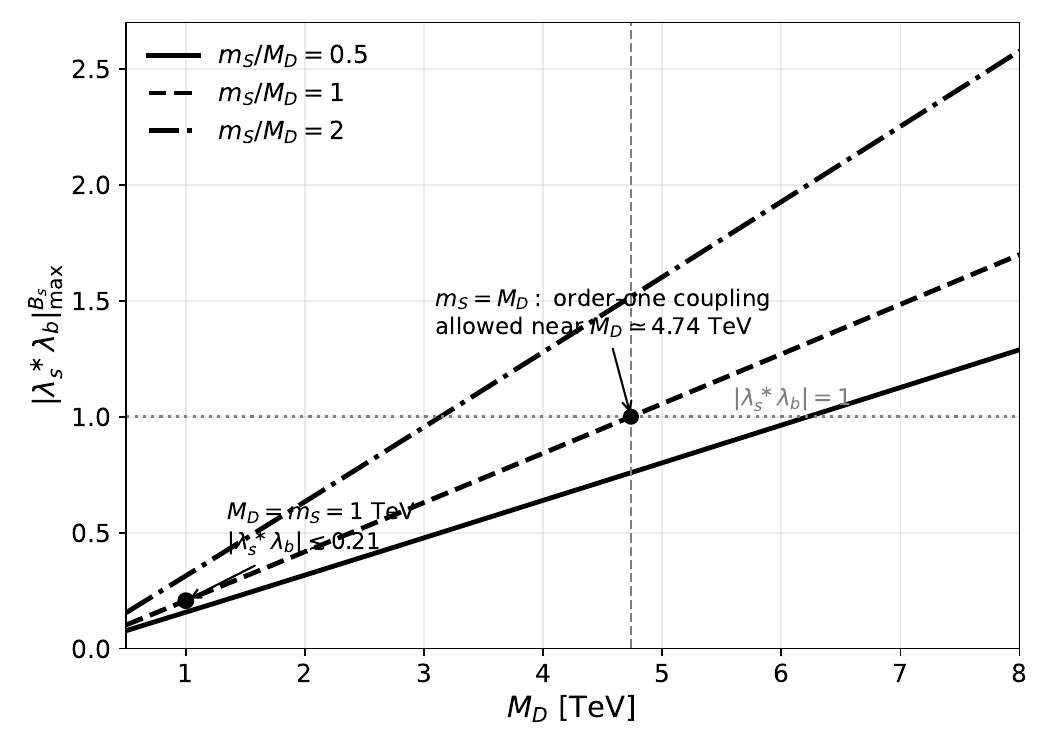}
\caption{Upper limit on $|\lambda_s^\ast\lambda_b|$ from $B_s-\bar B_s$ mixing using the conservative criterion $h_s<0.20$, shown for representative mass ratios $m_S/M_D=0.5,1,2$.}
\label{fig:lambdamax-bs}
\end{figure}

Using Eqs.~\eqref{eq:CRRbs-match-main},~\eqref{eq:Udb2-main}, and~\eqref{eq:hsdef}, then the condition from Eq.~\eqref{eq:deltabscriterion} giving
\begin{equation}
|\lambda_s^\ast\lambda_b|_{\rm max}^{B_s}
=
\left[
\frac{\delta_{B_s}\,|C_{LL}^{\rm SM}(\mu_b)|\,128\pi^2M_D^2}
{U_{\Delta B=2}\,G_{\rm box}(x)}
\right]^{1/2}.
\label{eq:lambdamax-bs}
\end{equation}
For the equal-mass case, $m_S=M_D$, we have $x=\frac{m_S^2}{M_D^2}=1,\,G_{\rm box}(1)=\frac{1}{3}$. After, substituting this into Eq.~\eqref{eq:lambdamax-bs}, and using
$\delta_{B_s}=0.20$, $|C_{LL}^{bs,\rm SM}(\mu_b)|\simeq
4.5\times10^{-11}~{\rm GeV}^{-2}$, and $U_{\Delta B=2}\simeq0.80$ for $M_D=1~{\rm TeV}$, gives
\begin{align}
|\lambda_s^\ast\lambda_b|_{\max}^{B_s}
=
\left[
\frac{
(0.20)(4.5\times10^{-11})(128\pi^2)(1000)^2
}
{
(0.80)(1/3)
}
\right]^{1/2}
\simeq 0.21 .
\end{align}
Therefore, for the equal-mass line, the mixing constraint can be written
approximately as
\begin{align}
|\lambda_s^\ast\lambda_b|_{\max}^{B_s}
\simeq
0.21
\left(\frac{M_D}{1~{\rm TeV}}\right)
\left(\frac{\delta_{B_s}}{0.20}\right)^{1/2}.
\end{align}
A stricter $10\%$ allowance gives $|\lambda_s^\ast\lambda_b|\lesssim0.15$, while a $5\%$ allowance gives $|\lambda_s^\ast\lambda_b|\lesssim0.10$ at $M_D=m_S=1~{\rm TeV}$. The numerical values for the conservative $20\%$ criterion are shown in Table~\ref{tab:bs-mixing-bound} and Figure~\ref{fig:lambdamax-bs}. The upper limit on the $|\lambda_s^\ast\lambda_b|$ from the $B_s-\bar B_s$  $h_s<0.20$, shown as a function of $M_D$ for mass ratios $m_S/M_D=0.5,1,2$. Since the mixing amplitude scales as $h_s\propto |\lambda_s^\ast\lambda_b|^2G_{\rm box}(x)/M_D^2$, the allowed coupling $|\lambda_s^\ast\lambda_b|$ increases approximately linearly with $M_D$. As we see that for larger $m_S/M_D$, the $G(x)$ is smaller, and hence the mixing constraint is weaker. The horizontal line marks $|\lambda_s^\ast\lambda_b|=1$.

\begin{table}[H]
\centering
\caption{Approximate upper bound on $|\lambda_s^\ast\lambda_b|$ from $B_s-\bar B_s$ mixing using $h_s<0.20$. The values include the leading-logarithmic running factor in Eq.~\eqref{eq:Udb2-main}.}
\label{tab:bs-mixing-bound}
\begin{tabular}{c c c c c}
\toprule
$m_S/M_D$ & $M_D=1~{\rm TeV}$ & $M_D=2~{\rm TeV}$ & $M_D=3~{\rm TeV}$ & $M_D=5~{\rm TeV}$ \\
\midrule
$0.5$ & $0.16$ & $0.32$ & $0.48$ & $0.80$ \\
$1.0$ & $0.21$ & $0.42$ & $0.63$ & $1.06$ \\
$2.0$ & $0.31$ & $0.63$ & $0.96$ & $1.60$ \\
\bottomrule
\end{tabular}
\end{table}

The numerical output along the equal mass line $m_S=M_D$ is given in Table~\ref{tab:mathematica-equalmass}. This table is useful because it shows, the relative importance of the $B\to X_s\gamma$ and $B_s-\bar B_s$ mixing constraints. The second and third columns show that the radiative correction rapidly falls below the percent level as the new particle masses increase. The fourth column then shows that the corresponding radiative only limit on $|\lambda_s^\ast\lambda_b|$ becomes very weak. By contrast, the fifth and sixth columns show that $B_s-\bar B_s$ mixing remains restrictive in the same mass range. The last column gives the combined upper limit, which is essentially identical to the $B_s$-mixing limit throughout the given range. As a result, the table provides the clearest numerical values for the main phenomenological conclusion: in the minimal benchmark, $B\to X_s\gamma$ is radiatively safe, where as $B_s-\bar B_s$ mixing controls the allowed $|\lambda_s^\ast\lambda_b|$.

\begin{table}[H]
\centering
\caption{Numerical values for the radiative bound uses $|C_{7\gamma}^{\rm NP,eff}(\mu_b)|<0.05$, while the mixing bound uses $h_s=|C_{RR}^{bs,\rm NP}/C_{LL}^{bs,\rm SM}|<0.20$. The final column gives the combined upper limit.}
\label{tab:mathematica-equalmass}
\resizebox{\textwidth}{!}{%
\begin{tabular}{c c c c c c c}
\toprule
$M_D=m_S$ [GeV]
& $|C_{7\gamma}^{\rm NP,eff}|$
& $100|C_{7\gamma}^{\rm NP,eff}/C_{7\gamma}^{\rm SM,eff}|$
& $|\lambda_s^\ast\lambda_b|_{\max}^{B\to X_s\gamma}$
& $h_s$ for $|\lambda_s^\ast\lambda_b|=1$
& $|\lambda_s^\ast\lambda_b|_{\max}^{B_s}$
& combined max \\
\midrule
$500$  & $5.49\times10^{-3}$ & $1.83$  & $9.11$    & $19.06$ & $0.102$ & $0.102$ \\
$750$  & $2.18\times10^{-3}$ & $0.727$ & $22.93$   & $8.37$  & $0.155$ & $0.155$ \\
$1000$ & $1.13\times10^{-3}$ & $0.377$ & $44.17$   & $4.67$  & $0.207$ & $0.207$ \\
$1500$ & $4.48\times10^{-4}$ & $0.149$ & $111.53$  & $2.05$  & $0.312$ & $0.312$ \\
$2000$ & $2.32\times10^{-4}$ & $0.0773$& $215.58$  & $1.15$  & $0.418$ & $0.418$ \\
$3000$ & $9.13\times10^{-5}$ & $0.0304$& $547.61$  & $0.504$ & $0.630$ & $0.630$ \\
$5000$ & $2.80\times10^{-5}$ & $0.00934$& $1784.91$ & $0.179$ & $1.056$ & $1.056$ \\
\bottomrule
\end{tabular}%
}
\end{table}

\begin{figure}[H]
\centering
\includegraphics[width=0.72\textwidth]{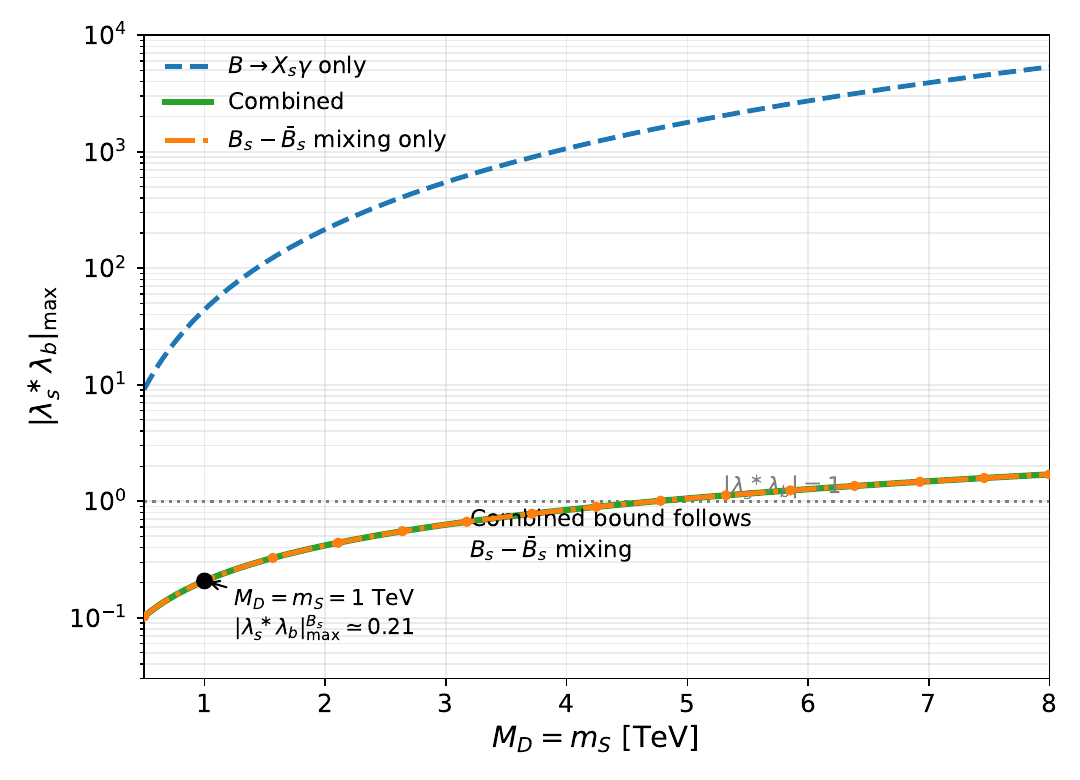}
\caption{Comparison of the radiative-only $B\to X_s\gamma$ bound, the
$B_s-\bar B_s$ mixing bound, and the combined upper limit on
$|\lambda_s^\ast\lambda_b|$.}
\label{fig:constraint-comparison}
\end{figure}

The Comparison of the $B\to X_s\gamma$ bound, the
$B_s-\bar B_s$ mixing bound, and the combined upper limit on
$|\lambda_s^\ast\lambda_b|$ along the equal-mass line $m_S=M_D$ is given in Figure~\ref{fig:constraint-comparison}.
The radiative bound from $B\to X_s\gamma$ is obtained using $|C_{7\gamma}^{\rm NP,eff}(\mu_b)|<0.05$, while the $B_s-\bar B_s$ mixing bound is obtained using $h_s<0.20$. The combined bound is taken as the stronger of these two constraints. Over the full mass range shown, the combined curve overlaps with the $B_s-\bar B_s$ mixing curve. This shows that $B_s-\bar B_s$ mixing gives the dominant constraint on the $|\lambda_s^\ast\lambda_b|$ in the minimal benchmark.

\begin{figure}[H]
\centering
\includegraphics[width=0.76\textwidth]{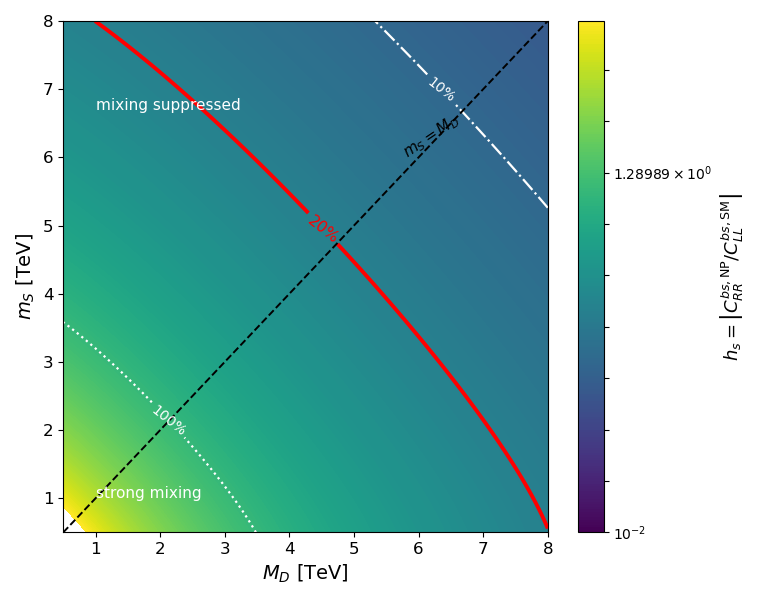}
\caption{2D scan of the $B_s-\bar B_s$ mixing ratio $h_s=|C_{RR}^{bs,\rm NP}/C_{LL}^{bs,\rm SM}|$ in the $(M_D,m_S)$ plane for $|\lambda_s^\ast\lambda_b|=1$. }
\label{fig:2dbs}
\end{figure}

In Figure~\ref{fig:2dbs}, we have the 2D scan of the $B_s-\bar B_s$ mixing ratio
$h_s=|C_{RR}^{bs,\rm NP}/C_{LL}^{bs,\rm SM}|$ in the $(M_D,m_S)$ plane for $|\lambda_s^\ast\lambda_b|=1$. The contour lines correspond to NP contributions of $5\%$, $10\%$, $20\%$, and$100\%$ relative to the SM mixing amplitude. The highlighted $20\%$ contour corresponds to the conservative condition $h_s<0.20$. The largest mixing effects occur when both new particles are relatively light, while increasing either $M_D$ or $m_S$ suppresses the new contribution.

\subsection{Simplified Flavor Likelihood}
\label{subsec:simplified-flavor-likelihood}

The $h_s<0.20$ condition is a useful way to show the dominant constraint in a transparent way. As a complementary check, we also construct a simplified flavor likelihood using the measured values of $\mathcal B( B\to X_s\gamma)$, $\Delta M_s$, and CP phase $\phi_s$. This likelihood is not meant to replace a full global CKM analysis. It is used only to check whether the cut-based result stays stable when we use the measured values and uncertainties directly.

We write the complex flavor product as
\begin{equation}
\lambda_s^\ast\lambda_b
=
\Lambda e^{i\varphi_\lambda},
\qquad
\Lambda=|\lambda_s^\ast\lambda_b|.
\label{eq:lambda-phase-likelihood}
\end{equation}
Then the radiative WCs scales as
\begin{equation}
C_{7\gamma}^{\rm NP,eff}(\mu_b)
=
\Lambda e^{i\varphi_\lambda}
\left.
C_{7\gamma}^{\rm NP,eff}(\mu_b)
\right|_{\Lambda=1},
\end{equation}
whereas the \(\Delta B=2\) coefficient scales as
\begin{equation}
C_{RR}^{bs,\rm NP}
\propto
\Lambda^2 e^{2i\varphi_\lambda}.
\end{equation}
For $\bar B\to X_s\gamma$, we use the same approximation given in Eq.~\eqref{eq:Rratio},
\begin{equation}
\mathcal B(\bar B\to X_s\gamma)_{\rm th}
=
\mathcal B(\bar B\to X_s\gamma)_{\rm SM}
\left|
1+
\frac{C_{7\gamma}^{\rm NP,eff}(\mu_b)}
{C_{7\gamma}^{\rm SM,eff}(\mu_b)}
\right|^2 .
\label{eq:bsgamma-likelihood-branching}
\end{equation}
The corresponding contribution to the simplified likelihood is
\begin{equation}
\chi^2_{B\to X_s\gamma}
=
\frac{
\left[
\mathcal B(\bar B\to X_s\gamma)_{\rm th}
-
\mathcal B(\bar B\to X_s\gamma)_{\rm exp}
\right]^2
}
{
\sigma_{B\to X_s\gamma,{\rm exp}}^2
+
\sigma_{B\to X_s\gamma,{\rm SM}}^2
}.
\label{eq:chi2-bsgamma-likelihood}
\end{equation}
Numerically, we use
\begin{equation}
\mathcal B(\bar B\to X_s\gamma)_{\rm exp}
=
(3.49\pm0.19)\times10^{-4},
\qquad
\mathcal B(\bar B\to X_s\gamma)_{\rm SM}
=
(3.40\pm0.17)\times10^{-4},
\label{eq:likelihood-bsgamma-inputs}
\end{equation}
as given in Eqs.~\eqref{eq:hflav} and \eqref{eq:smbr}.

For $B_s-\bar B_s$ mixing in the complex-scalar benchmark we parameterize the full amplitude as
\begin{equation}
M_{12}^s
=
M_{12}^{s,\rm SM}
\left(1+h_s e^{2i\varphi_\lambda}\right),
\qquad
h_s
=
\left|
\frac{C_{RR}^{bs,\rm NP}}
{C_{LL}^{bs,\rm SM}}
\right|.
\label{eq:m12-likelihood}
\end{equation}
This gives
\begin{equation}
\Delta M_s^{\rm th}
=
\Delta M_s^{\rm SM}
\left|1+h_s e^{2i\varphi_\lambda}\right|,
\label{eq:delta-ms-likelihood}
\end{equation}
and the CP phase is shifted according to
\begin{equation}
\phi_s^{\rm th}
=
\phi_s^{\rm SM}
+
\arg\left(1+h_s e^{2i\varphi_\lambda}\right).
\label{eq:phis-likelihood}
\end{equation}
The corresponding likelihood contributions are
\begin{equation}
\chi^2_{\Delta M_s}
=
\frac{
\left[
\Delta M_s^{\rm th}
-
\Delta M_s^{\rm exp}
\right]^2
}
{
\sigma_{\Delta M_s,{\rm exp}}^2
+
\sigma_{\Delta M_s,{\rm SM}}^2
},
\label{eq:chi2-deltams-likelihood}
\end{equation}
and
\begin{equation}
\chi^2_{\phi_s}
=
\frac{
\left[
\phi_s^{\rm th}
-
\phi_s^{\rm exp}
\right]^2
}
{
\sigma_{\phi_s,{\rm exp}}^2
+
\sigma_{\phi_s,{\rm SM}}^2
}.
\label{eq:chi2-phis-likelihood}
\end{equation}
For the mixing inputs we use
\begin{equation}
\Delta M_s^{\rm exp}=17.766\pm0.006~{\rm ps}^{-1},
\qquad
\Delta M_s^{\rm SM}=18.4\pm1.0~{\rm ps}^{-1},
\label{eq:likelihood-deltams-inputs}
\end{equation}
where the SM uncertainty is a symmetrized version of the value quoted in
Ref.~\cite{DiLuzio:2019jyq}. For the mixing phase we use
\begin{equation}
\phi_s^{\rm exp}=-0.052\pm0.013~{\rm rad}\text{\cite{HeavyFlavorAveragingGroupHFLAV:2024ctg}},
\qquad
\phi_s^{\rm SM}=-0.0365\pm0.0013~{\rm rad}\text{\cite{LHCb:2021wte}}.
\label{eq:likelihood-phis-inputs}
\end{equation}
The total simplified flavor likelihood is then
\begin{equation}
\chi^2_{\rm flavor}
=
\chi^2_{B\to X_s\gamma}
+
\chi^2_{\Delta M_s}
+
\chi^2_{\phi_s}.
\label{eq:chi2-flavor-total}
\end{equation}

\begin{figure}[H]
\centering
\includegraphics[width=0.70\textwidth]{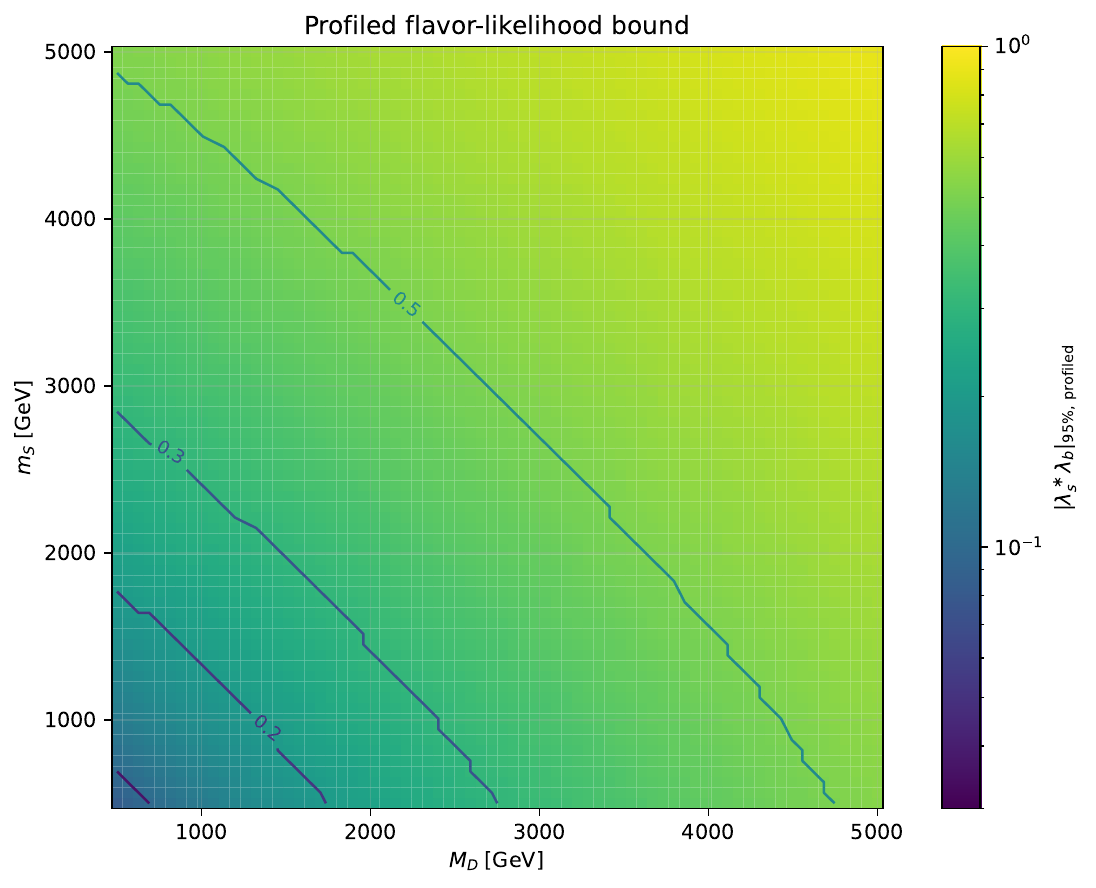}
\caption{Profiled \(95\%\) upper limit on
\(\Lambda=|\lambda_s^\ast\lambda_b|\) in the \((M_D,m_S)\) plane from the
simplified flavor likelihood based on
\(\mathcal B(\bar B\to X_s\gamma)\), \(\Delta M_s\), and \(\phi_s\).
The phase \(\varphi_\lambda\) is profiled over, and the conservative
perturbativity condition \(\Lambda\leq1\) is imposed.}
\label{fig:realistic-lambda95-profiled}
\end{figure}

At each point in the \((M_D,m_S)\) plane we scan over
\(\Lambda\leq 1\) and \(\varphi_\lambda\in[-\pi,\pi]\). The condition
\(\Lambda\leq1\) corresponds to the conservative perturbativity choice
\(|\lambda_s|,|\lambda_b|<1\). After profiling over the phase
\(\varphi_\lambda\), we define the approximate profiled \(95\%\) upper limit on
\(\Lambda\) by
\begin{equation}
\Delta\chi^2_{\rm flavor}(\Lambda)
=
\chi^2_{\rm flavor}(\Lambda)
-
\chi^2_{\rm flavor,min}
\leq 3.84,
\label{eq:profiled-95-criterion}
\end{equation}
which is the usual one-parameter \(95\%\) criterion. Here ``profiling over $\varphi_\lambda$'' means that, for each fixed value
of the coupling magnitude $\Lambda$, we minimize
$\chi^2_{\rm flavor}(\Lambda,\varphi_\lambda)$ with respect to the phase
$\varphi_\lambda$. Thus the resulting bound on $\Lambda$ corresponds to the
best possible choice of the CP-violating phase at each value of
$\Lambda$.  
The profiled bound is shown in Figure~\ref{fig:realistic-lambda95-profiled}. The likelihood-based gives the same main conclusion as the conservative mixing criterion. The allowed value of $|\lambda_s^\ast\lambda_b|$ is mainly controlled by $B_s-\bar B_s$ mixing in the complex-scalar benchmark, while $B\to X_s\gamma$ remains a useful but weaker consistency check.

\begin{figure}[H]
\centering
\includegraphics[width=0.70\textwidth]{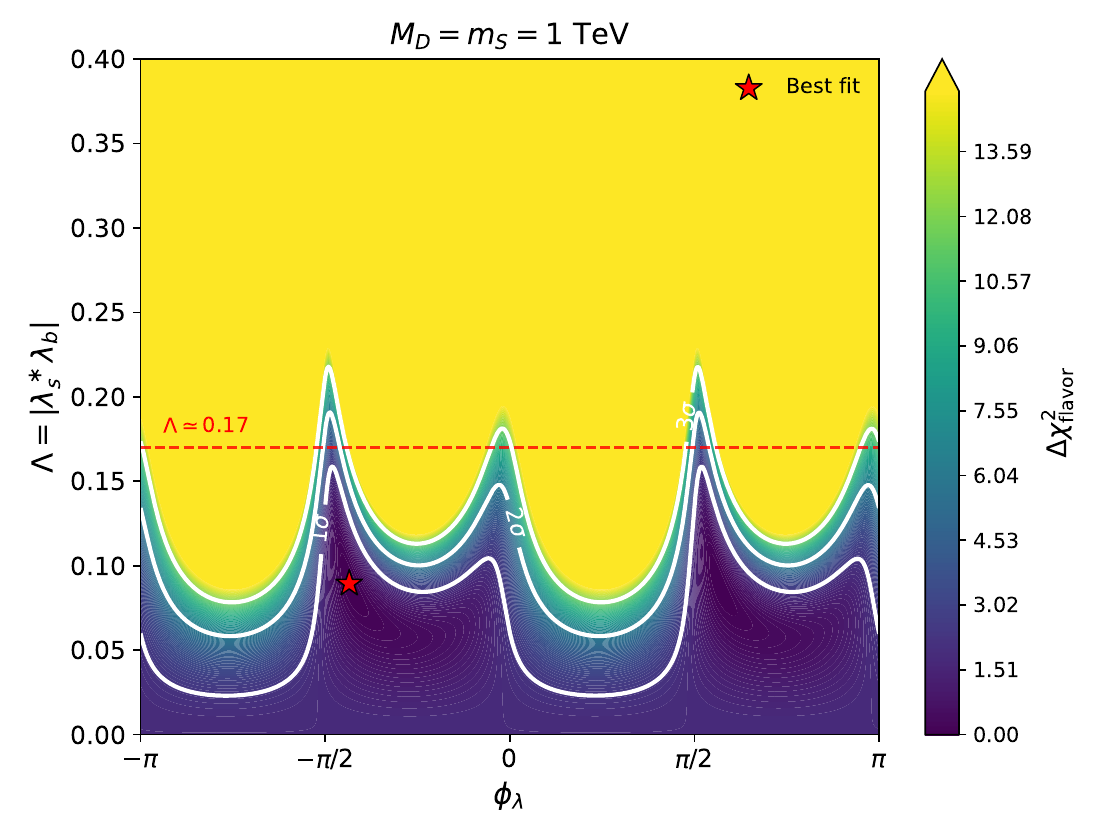}
\caption{ Complex-scalar benchmark likelihood contours in the
\((\Lambda,\varphi_\lambda)\) plane for \(M_D=m_S=1~{\rm TeV}\). The contours
correspond to \(\Delta\chi^2=2.30,\,6.18,\,11.83\), approximately representing
the usual \(1\sigma\), \(2\sigma\), and \(3\sigma\) regions for two scanned
parameters.}
\label{fig:realistic-benchmark-deltachi2}
\end{figure}

In Figure~\ref{fig:realistic-benchmark-deltachi2}, we plot the contours of $\Delta\chi^2_{\rm flavor}$ in the $(\Lambda,\phi_\lambda)$ plane for the complex scalar benchmark point $M_D=m_S=1~{\rm TeV}$, where $\Lambda=|\lambda_s^\ast\lambda_b|$ and $\lambda_s^\ast\lambda_b=\Lambda e^{i\phi_\lambda}$. The contour levels correspond to $\Delta\chi^2=2.30,6.18,11.83$, approximately representing the usual $1\sigma$, $2\sigma$, and $3\sigma$ regions for two scanned parameters. The best-fit point is shown by the star. Furhter, the figure show the phase dependence of the simplified flavor likelihood and also shows that the allowed region favors a small $|\lambda_s^\ast\lambda_b|$ coupling, mainly due to the strong $B_s-\bar B_s$ mixing constraint.

The $M_D=m_S$ benchmark values obtained from this likelihood are summarized in
Table~\ref{tab:realistic-likelihood-equalmass}. At
\(M_D=m_S=1~{\rm TeV}\), the profiled likelihood gives $|\lambda_s^\ast\lambda_b|_{95\%}\simeq0.17$, which is close to but slightly stronger than the \(h_s<0.20\) value. The agreement between the two
methods shows that the simple mixing criterion captures the dominant physical
effect, while the likelihood treatment provides a more realistic visualization
of the allowed parameter space.

\begin{table}[H]
\centering
\caption{The $M_D=m_S$ benchmark values from the simplified
profiled flavor likelihood. Here $\Lambda=|\lambda_s^\ast\lambda_b|$, and the phase $\varphi_\lambda$ is profiled over.}
\label{tab:realistic-likelihood-equalmass}
\renewcommand{\arraystretch}{1.2}
\begin{tabular}{c c c c}
\toprule
\(M_D=m_S\) [GeV]
&
\(|C_{7\gamma}^{\rm NP,eff}(\mu_b)|_{\Lambda=1}\)
&
\(h_s|_{\Lambda=1}\)
&
\(\Lambda_{95\%,{\rm profiled}}\)
\\
\midrule
1000 & \(1.13\times10^{-3}\) & \(4.67\) & \(0.172\) \\
1500 & \(4.48\times10^{-4}\) & \(2.05\) & \(0.261\) \\
2000 & \(2.32\times10^{-4}\) & \(1.15\) & \(0.350\) \\
\bottomrule
\end{tabular}
\end{table}

\section{Conclusion}
\label{sec:conclusion}

In this work, we studied a minimal extension of the SM containing a neutral gauge-singlet scalar ${\cal S}$ and a vector-like down-type quark $D$. Here, the ${\cal S}$  is catagorized into two benchmarks: a real singlet scalar $S_R=S_R^\dagger$ and a complex singlet scalar $\Phi\neq\Phi^\dagger$. The new interaction $\lambda_i {\cal S}\bar D_L d_{Ri}$ generates FCNC effects through the coupling product $\lambda_s^\ast\lambda_b$. We focused on two important flavor observables: the inclusive radiative decay $B\to X_s\gamma$ and $B_s-\bar B_s$ mixing.
The scalar is neutral under electromagnetism and $SU(3)_c$. Therefore, the photon $\gamma$ and gluon $g$ can only be emitted from the internal $D$ quark line. This gives the matching relation ${C_{7\gamma}^{\rm NP}}/{C_{8G}^{\rm NP}}=Q_D=-{1}/{3}$.

We computed the one-loop matching contributions to the electromagnetic $O_{7\gamma}$ and
chromomagnetic $O_{8G}$ dipole operators and included the leading-logarithmic QCD
evolution down to the scale $\mu_b$. The contribution to $C_{7\gamma}^{\rm eff}(\mu_b)$ remains small in a wide region of the TeV-scale parameter space. In particular, in the equal-mass case $M_D=m_S=1~{\rm TeV}$ with $|\lambda_s^\ast\lambda_b|=1$,leading-logarithmic running gives $|C_{7\gamma}^{\rm NP,eff}(\mu_b)|\simeq1.1\times10^{-3}$, which is only about $0.4\%$ of $|C_{7\gamma}^{\rm SM,eff}(\mu_b)|\simeq0.30$. Thus, $B\to X_s\gamma$ alone does not strongly constrain the minimal non-chirality enhanced benchmarks unless we have that the new particles are lighter, the flavor $|\lambda_s^\ast\lambda_b|$ is much larger, or chirality enhanced interactions are introduced.

For the case of $\Delta B=2$ the two scalar choices differ. For ${\cal S}=S_R$ benchmark, the direct box is accompanied by a crossed box Feynman diagram. The crossed box has the same loop function and color factor as the direct box but with opposite negative sign, Consequently,
$C_{RR}^{bs,{\rm full}}(\mu_D)=0$ at leading dimension-six operator. As a result, the direct box Feynman diagram $B_s-\bar B_s$ strong bound does not apply to the exact minimal real-scalar benchmark. For the complex-scalar ${\cal S}=\Phi$ benchmark, the real-scalar ${S_R}$ crossed contraction is absent, where the amplitude scale as $C_{RR}^{bs,\rm NP}\propto [{(\lambda_s^\ast\lambda_b)^2}/{M_D^2}]G_{\rm box}\left({m_S^2}/{M_D^2}\right)$. This dependence is quadratic in $|\lambda_s^\ast\lambda_b|$, so $B_s-\bar B_s$ mixing gives a more stronger constraint on the flavor-coupling $|\lambda_s^\ast\lambda_b|$. For $M_D=m_S=1~{\rm TeV}$, an $\mathcal{O}(1)$ coupling produces a new contribution to $B_s-\bar B_s$ mixing that is larger than the conservative allowed range. It require the new contribution to be less than $20\%$ of the SM one, we find approximately $|\lambda_s^\ast\lambda_b|\lesssim 0.21$ for the equal-mass TeV scale.

We also include a simplified likelihood analysis based on $\mathcal B(B\to X_s\gamma)$, $\Delta M_s$, and $\phi_s$. In this analysis, the $\lambda_s^\ast\lambda_b$ is allowed to be complex,
$\lambda_s^\ast\lambda_b=\Lambda e^{i\varphi_\lambda}$, and profiled over the phase $\varphi_\lambda$. The result support the same conclusion as the simpler cut-based approach. The allowed parameter space is mainly limited by $B_s-\bar B_s$ mixing, while $B\to X_s\gamma$ gives a weaker but still useful complementary check. For the TeV-scale complex-scalar benchmark of equal-mass, the profiled likelihood flavor gives an upper bound on the coupling product
$(\lambda_s^\ast\lambda_b)$ at the level of a few tenths. This is consistent
with the simpler conservative mixing criterion $h_s<0.20$. This confirms that $B_s-\bar B_s$ mixing provides the dominant constraint in complex scalar benchmark.
The main conclusion is therefore not simply that $B_s-\bar B_s$ mixing always dominates. Rather, the scalar nature matters: in the real scalar benchmark, the leading direct box and crossed-box Feynman diagrams combine in such a way that they cancel each other. while the complex-scalar benchmark is controlled primarily by $B_s-\bar B_s$ mixing.  This shows that $B_s-\bar B_s$ mixing provides the leading flavor constraint in the complex scalar benchmark, while $B\to X_s\gamma$ serves as a clean complementary probe of the dipole structure.

\clearpage
\appendix

\section*{Appendix}

\section{Loop function $F(x)$ from the Feynman-parameter integral}\label{sec:appendixA}

For the calculation of Eq.~\eqref{eq:fullamp_schematic}, we combine the denominators, shift the loop momentum, and project the result onto the dipole structure. The dimensionless loop function can then be written as
\begin{equation}
F(x)=
\frac12\int_0^1 dz\,
\frac{z^2(1-z)}{z+x(1-z)},
\qquad
x=\frac{m_S^2}{M_D^2}.
\label{eq:F-integral}
\end{equation}
The denominator can be rewritten as
\begin{equation*}
z+x(1-z)=x+(1-x)z.
\end{equation*}
Define
\begin{equation*}
t=x+(1-x)z,
\qquad
z=\frac{t-x}{1-x},
\qquad
1-z=\frac{1-t}{1-x},
\qquad
 dz=\frac{dt}{1-x}.
\end{equation*}
When $z=0$, $t=x$, and when $z=1$, $t=1$.  Therefore
\begin{align}
\int_0^1 dz\,
\frac{z^2(1-z)}{x+(1-x)z}
&=
\frac{1}{(1-x)^4}
\int_x^1 dt\,
\frac{(t-x)^2(1-t)}{t}
\nonumber\\
&=
\frac{1}{(1-x)^4}
\int_x^1 dt\,
\left[-t^2+(1+2x)t-(2x+x^2)+\frac{x^2}{t}\right]\nonumber\\
\int_0^1 dz\,
\frac{z^2(1-z)}{x+(1-x)z}
&=
\frac{
\frac16-x+\frac{x^2}{2}+\frac{x^3}{3}-x^2\ln x
}{(1-x)^4}.
\end{align}\label{eq:F-integral-expanded}
Multiplying by the prefactor $1/2$ in Eq.~\eqref{eq:F-integral}, give
\begin{equation}
F(x)=
\frac{1-6x+3x^2+2x^3-6x^2\ln x}{12(1-x)^4}.
\label{eq:F-closed}
\end{equation}
The Mathematica check gives
\begin{equation}
F_{\rm integral}(x)-F_{\rm closed}(x)=0,
\end{equation}
for $x>0$ and $x\neq1$. The limiting values are
\begin{equation}
\lim_{x\to1}F(x)=\frac{1}{24},
\qquad
\lim_{x\to0^+}F(x)=\frac{1}{12}.
\end{equation}
These checks are useful for verifying the closed-form expression used in the matching calculation. 

\section{Details of the Matching Calculation}\label{sec:appendixB}
\subsection*{Matching for the electromagnetic coefficient}
The dipole part of the full photon amplitude is parameterized as
\begin{equation}
\mathcal M_\gamma^{\rm full}
=
\frac{e}{16\pi^2}m_b A_7^{\rm full}
(\bar s\sigma^{\mu\nu}P_Rb)F_{\mu\nu}.
\label{eq:full-dipole-photon}
\end{equation}
Similarly, the full chromomagnetic amplitude is written as
\begin{equation}
\mathcal M_g^{\rm full}
=
\frac{g_s}{16\pi^2}m_b A_8^{\rm full}
(\bar s\sigma^{\mu\nu}T^aP_Rb)G^a_{\mu\nu}.
\label{eq:full-dipole-gluon}
\end{equation}
Here $A_7^{\rm full}$ and $A_8^{\rm full}$ are dimensionful coefficients of order $1/M_D^2$. 
\begin{align}
A_7^{\rm full}
&=
\xi_A\frac{\lambda_s^\ast\lambda_b}{2M_D^2}Q_DF(x),
\label{eq:A7-full-corrected}
\\
A_8^{\rm full}
&=
\xi_A\frac{\lambda_s^\ast\lambda_b}{2M_D^2}F(x).
\label{eq:A8-full-corrected}
\end{align}
The new-physics part of the effective Hamiltonian is taken as
\begin{equation}
\mathcal H_{\rm eff}^{\rm NP}
=
-
\frac{4G_F}{\sqrt2}V_{tb}V_{ts}^\ast
\left[
C_{7\gamma}^{\rm NP}O_{7\gamma}
+
C_{8G}^{\rm NP}O_{8G}
\right].
\label{eq:Heff}
\end{equation}
The dipole operators are
\begin{align}
O_{7\gamma}
&=
\frac{e}{16\pi^2}m_b
(\bar s\sigma^{\mu\nu}P_Rb)F_{\mu\nu},
\label{eq:O7}
\\
O_{8G}
&=
\frac{g_s}{16\pi^2}m_b
(\bar s\sigma^{\mu\nu}T^aP_Rb)G^a_{\mu\nu}.
\label{eq:O8}
\end{align}
The photon part of Eq.~\eqref{eq:Heff} is
\begin{equation}
\mathcal H_7^{\rm NP}
=
-
\frac{4G_F}{\sqrt2}V_{tb}V_{ts}^\ast
C_{7\gamma}^{\rm NP}O_{7\gamma}.
\end{equation}
Substituting Eq.~\eqref{eq:O7}, the effective-theory amplitude is
\begin{equation}
\mathcal M_\gamma^{\rm eff}
=
-
\frac{4G_F}{\sqrt2}V_{tb}V_{ts}^\ast
C_{7\gamma}^{\rm NP}
\frac{e}{16\pi^2}m_b
(\bar s\sigma^{\mu\nu}P_Rb)F_{\mu\nu}.
\label{eq:Meff-photon}
\end{equation}
The full-theory amplitude is Eq.~\eqref{eq:full-dipole-photon}.  Matching requires
\begin{equation}
\mathcal M_\gamma^{\rm full}=\mathcal M_\gamma^{\rm eff}.
\end{equation}
Thus
\begin{align}
&\frac{e}{16\pi^2}m_b A_7^{\rm full}
(\bar s\sigma^{\mu\nu}P_Rb)F_{\mu\nu}
\nonumber\\
&\hspace{2cm}=
-
\frac{4G_F}{\sqrt2}V_{tb}V_{ts}^\ast
C_{7\gamma}^{\rm NP}
\frac{e}{16\pi^2}m_b
(\bar s\sigma^{\mu\nu}P_Rb)F_{\mu\nu}.
\end{align}
We obtains
\begin{equation}
A_7^{\rm full}
=
-
\frac{4G_F}{\sqrt2}V_{tb}V_{ts}^\ast
C_{7\gamma}^{\rm NP}.
\label{eq:A7-match}
\end{equation}
Solving for $C_{7\gamma}^{\rm NP}$ gives
\begin{equation}
C_{7\gamma}^{\rm NP}
=
-
\frac{\sqrt2}{4G_FV_{tb}V_{ts}^\ast}
A_7^{\rm full}
=
-
\frac{1}{2\sqrt2\,G_FV_{tb}V_{ts}^\ast}
A_7^{\rm full}.
\label{eq:C7-in-terms-of-A7}
\end{equation}
Now substitute Eq.~\eqref{eq:A7-full-corrected}:
\begin{align*}
C_{7\gamma}^{\rm NP}(\mu_D)
&=
-
\frac{1}{2\sqrt2\,G_FV_{tb}V_{ts}^\ast}
\left[
\xi_A\frac{\lambda_s^\ast\lambda_b}{2M_D^2}Q_DF(x)
\right]
\\
&=
-
\xi_A
\frac{\lambda_s^\ast\lambda_b}{4\sqrt2\,G_FV_{tb}V_{ts}^\ast M_D^2}
Q_DF(x).
\end{align*}
Defining the convention-dependent sign $\xi=-\xi_A$, we obtain
\begin{equation}
C_{7\gamma}^{\rm NP}(\mu_D)
=
\xi
\frac{\lambda_s^\ast\lambda_b}{4\sqrt2\,G_FV_{tb}V_{ts}^\ast M_D^2}
Q_DF(x).
\label{eq:C7-final}
\end{equation}

\subsection*{Matching for the chromomagnetic coefficient}

The same steps apply to the gluon amplitude.  From Eq.~\eqref{eq:Heff}, the chromomagnetic part is
\begin{equation}
\mathcal H_8^{\rm NP}
=
-
\frac{4G_F}{\sqrt2}V_{tb}V_{ts}^\ast
C_{8G}^{\rm NP}O_{8G}.
\end{equation}
Substituting Eq.~\eqref{eq:O8}, the effective-theory amplitude is
\begin{equation}
\mathcal M_g^{\rm eff}
=
-
\frac{4G_F}{\sqrt2}V_{tb}V_{ts}^\ast
C_{8G}^{\rm NP}
\frac{g_s}{16\pi^2}m_b
(\bar s\sigma^{\mu\nu}T^aP_Rb)G^a_{\mu\nu}.
\label{eq:Meff-gluon}
\end{equation}
Matching Eq.~\eqref{eq:full-dipole-gluon} to Eq.~\eqref{eq:Meff-gluon} gives
\begin{equation*}
A_8^{\rm full}
=
-
\frac{4G_F}{\sqrt2}V_{tb}V_{ts}^\ast
C_{8G}^{\rm NP}.
\end{equation*}
Hence
\begin{equation*}
C_{8G}^{\rm NP}
=
-
\frac{1}{2\sqrt2\,G_FV_{tb}V_{ts}^\ast}
A_8^{\rm full}.
\end{equation*}
Using Eq.~\eqref{eq:A8-full-corrected},
\begin{align}
C_{8G}^{\rm NP}(\mu_D)
&=
-
\frac{1}{2\sqrt2\,G_FV_{tb}V_{ts}^\ast}
\left[
\xi_A\frac{\lambda_s^\ast\lambda_b}{2M_D^2}F(x)
\right]
\nonumber\\
&=
-
\xi_A
\frac{\lambda_s^\ast\lambda_b}{4\sqrt2\,G_FV_{tb}V_{ts}^\ast M_D^2}
F(x).
\end{align}
Again defining $\xi=-\xi_A$, we finds
\begin{equation}
C_{8G}^{\rm NP}(\mu_D)
=
\xi
\frac{\lambda_s^\ast\lambda_b}{4\sqrt2\,G_FV_{tb}V_{ts}^\ast M_D^2}
F(x).
\label{eq:C8-final}
\end{equation}

\subsection*{Ratio of electromagnetic and chromomagnetic coefficients}

Dividing Eq.~\eqref{eq:C7-final} by Eq.~\eqref{eq:C8-final}, all common factors cancel:
\begin{align*}
\frac{C_{7\gamma}^{\rm NP}}{C_{8G}^{\rm NP}}
&=
\frac{
\xi
\dfrac{\lambda_s^\ast\lambda_b}{4\sqrt2\,G_FV_{tb}V_{ts}^\ast M_D^2}
Q_DF(x)
}{
\xi
\dfrac{\lambda_s^\ast\lambda_b}{4\sqrt2\,G_FV_{tb}V_{ts}^\ast M_D^2}
F(x)
}
\\
&=Q_D.
\end{align*}
Since the vector-like quark has electric charge $Q_D=-1/3$,
the final relation is
\begin{equation}
\frac{C_{7\gamma}^{\rm NP}}{C_{8G}^{\rm NP}}
=
Q_D
=-\frac13.
\label{eq:ratio-final}
\end{equation}

\section{Details of the $B_s-\bar B_s$ Box Contribution}\label{sec:appendixC}
\label{app:bs-mixing}

This appendix gives the mathematical details behind the $B_s-\bar B_s$ box discussion in Section~\ref{sec:mixing}. The interaction is
\begin{equation}
\mathcal L_{\rm int}=-\lambda_i{\cal S}\bar D_Ld_{Ri}+{\rm h.c.}
=-\lambda_i{\cal S}\bar D P_Rd_i-\lambda_i^\ast{\cal S}^\dagger\bar d_iP_LD.
\end{equation}
The relevant right-handed operator is
\begin{equation}
O_{RR}^{bs}
=
(\bar s_\alpha\gamma_\mu P_Rb_\alpha)
(\bar s_\beta\gamma^\mu P_Rb_\beta).
\end{equation}
For the direct scalar--fermion box, the numerator contains
\begin{equation}
P_L(\slashed k+M_D)P_R=P_L\slashed k P_R,
\end{equation}
so the internal mass term does not generate an $M_D/m_b$ enhancement. After loop integration, $k_\mu k_\nu\to g_{\mu\nu}k^2/d$, giving the vector right-handed operator above. The loop integral entering the coefficient has the generic form
\begin{equation}
I_{\rm box}
\propto
\int\frac{d^4k}{(2\pi)^4}
\frac{k^2}{(k^2-M_D^2)^2(k^2-m_{\cal S}^2)^2}.
\end{equation}
Combining denominators with a Feynman parameter gives
\begin{equation}
\frac{1}{(k^2-M_D^2)^2(k^2-m_{\cal S}^2)^2}
=
6\int_0^1 dz\,
\frac{z(1-z)}{[k^2-zM_D^2-(1-z)m_{\cal S}^2]^4}.
\end{equation}
Using
\begin{equation}
\int\frac{d^4k}{(2\pi)^4}
\frac{k^2}{(k^2-\Delta)^4}
=\frac{i}{16\pi^2}\frac{1}{3\Delta},
\end{equation}
one obtains the dimensionless function
\begin{equation}
G_{\rm box}(x)
=
2\int_0^1dz\,
\frac{z(1-z)}{z+x(1-z)},
\qquad
x=\frac{m_{\cal S}^2}{M_D^2}.
\end{equation}
The closed form is
\begin{equation}
G_{\rm box}(x)
=
\frac{1-x^2+2x\ln x}{(1-x)^3},
\qquad
G_{\rm box}(1)=\frac13,
\qquad
G_{\rm box}(0)=1.
\end{equation}
Including the spin, color, and symmetry factors in the operator normalization used in Section~\ref{sec:matching}, the direct box matching coefficient is
\begin{equation}
C_{RR}^{bs,{\rm dir}}(\mu_D)
=
\frac{(\lambda_s^\ast\lambda_b)^2}{128\pi^2M_D^2}
G_{\rm box}(x).
\end{equation}
For a real scalar, ${\cal S}=S_R=S_R^\dagger$, the scalar lines do not carry a conserved scalar-flow arrow. Therefore the ordinary box is accompanied by a crossed scalar-box topology. The generic real-scalar crossed-box result is obtained from the direct Feynman loop by interchanging scalar labels and replacing the relevant box loop function by $G\to -G$~\cite{Arnan:2019uhr}. In the minimal model with only one scalar and one vector-like quark, this interchange leaves the masses, coupling product, and color factor unchanged. Hence
\begin{equation}
C_{RR}^{bs,{\rm cross}}(\mu_D)
=
-
\frac{[(\lambda_s^R)^\ast\lambda_b^R]^2}{128\pi^2M_D^2}
G_{\rm box}(x),
\end{equation}
and
\begin{equation}
C_{RR}^{bs,{\rm full}}(\mu_D)
=
C_{RR}^{bs,{\rm dir}}(\mu_D)+C_{RR}^{bs,{\rm cross}}(\mu_D)=0.
\end{equation}
For a complex scalar, ${\cal S}=\Phi\neq\Phi^\dagger$, the scalar propagator preserves the scalar flow and the real-scalar crossed contraction is absent. The leading complex-scalar coefficient is therefore
\begin{equation}
C_{RR}^{bs}(\mu_D)
=
\frac{[\lambda_s^\ast\lambda_b]^2}{128\pi^2M_D^2}
G_{\rm box}(x),
\qquad x=\frac{m_S^2}{M_D^2}.
\end{equation}

%\bibliographystyle{myJHEP}
%\bibliography{References}

\providecommand{\url}[1]{#1}\providecommand{\href}[2]{#2}\begingroup\raggedright\endgroup

\end{document}